\documentstyle[emulateapj,psfig,apjfonts]{article}

\submitted{Version: \today ~~ 
Received 2004 07 24; Accepted: 2004 08 17}

\def\gsim{\mathrel{\raise0.35ex\hbox{$\scriptstyle >$}\kern-0.6em
\lower0.40ex\hbox{{$\scriptstyle \sim$}}}}
\def\lsim{\mathrel{\raise0.35ex\hbox{$\scriptstyle <$}\kern-0.6em
\lower0.40ex\hbox{{$\scriptstyle \sim$}}}}
\def\gs{\mathrel{\raise0.35ex\hbox{$\scriptstyle >$}\kern-0.6em
\lower0.40ex\hbox{{$\scriptstyle \sim$}}}}
\def\ls{\mathrel{\raise0.35ex\hbox{$\scriptstyle <$}\kern-0.6em
\lower0.40ex\hbox{{$\scriptstyle \sim$}}}}

\def\kms{\,\hbox{km}\,\hbox{s}^{-1}}

\def\Wm2{\,\hbox{W}\,\hbox{m}^{-2}}

\makeatother

\lefthead{Swinbank et al.}
\righthead{Rest-frame Optical Spectra of SCUBA Galaxies}

\begin{document}

\title
{The Rest-frame Optical Spectra of SCUBA Galaxies}

\author{
A.\,M.\ Swinbank,\altaffilmark{1}
Ian Smail,\altaffilmark{1}
S.\,C.\ Chapman,\altaffilmark{2}
A.\,W.\ Blain,\altaffilmark{2}
R.\,J.\ Ivison\altaffilmark{3,4}
\& W.\,C.\ Keel\altaffilmark{5}
}

\setcounter{footnote}{0}

\altaffiltext{1}{Institute for
Computational Cosmology, Department of Physics, University of Durham, South
  Road, Durham DH1 3LE, UK -- Email: a.m.swinbank@dur.ac.uk}
\altaffiltext{2}{Astronomy Department, California Institute of
  Technology, 105-24, Pasadena, CA 91125, USA}
\altaffiltext{3}{Astronomy Technology Centre, Royal Observatory,
  Blackford Hill, Edinburgh, EH19 3HJ, UK}
\altaffiltext{4} {Institute for Astronomy, University of Edinburgh,
Edinburgh,EH19 3HJ, UK}
\altaffiltext{5}{Department of Physics and Astronomy, University of
  Alabama, Tuscaloosa, AL 35487, USA}
\setcounter{footnote}{4}

\begin{abstract}
  We present near-infrared spectroscopy and narrow-band imaging at the
  wavelength of redshifted H$\alpha$ for a sample of 30 high-redshift,
  far-infrared luminous galaxies.  This sample is selected from surveys
  in the sub-millimeter, millimeter and radio wavebands and has
  complete redshift coverage with a median redshift of $z\sim2.4$.  We
  use our data to measure the H$\alpha$ properties of these systems and
  to gauge the prevalence of active galactic nuclei (AGN) in these
  galaxies through their [N{\sc ii}]/H$\alpha$ ratios and H$\alpha$
  line widths.  Removing obvious AGN, we find that the predicted
  H$\alpha$ star formation rates in this diverse population are
  suppressed (by a factor of $\sim10$) compared to those derived from
  their far-infrared luminosities.  Using the AGN indicators provided
  by our near-infrared spectra, we estimate that AGN are present in at
  least 40\% of the galaxies in our sample. To further investigate
  this, we construct a composite rest-frame spectrum for both the
  entire sample and for those galaxies which individually show no signs
  of nuclear activity. We find [N{\sc ii}]/H$\alpha$ ratios for both
  composite spectra which suggest that the energy output of the
  galaxies is star-formation- rather than AGN-dominated.  However, we
  also find that the H$\alpha$ line in the composite non-AGN spectrum
  is best fit with an underlying broad-line component with a
  narrow/broad flux ratio of $0.45\pm0.20$.  The median H$\alpha$ line
  width for our sample (removing obvious AGN) is
  $400\pm70$\,km\,s$^{-1}$ (FWHM), and the typical spatial extent of
  the H$\alpha$ emission in our narrow-band observations is $\ls
  4$--8\,kpc, which indicates a dynamical mass of
  1--2$\times10^{11}M_{\odot}$ with corresponding dynamical times of
  10--20\,Myr.  Using both high-resolution imaging and
  spectroscopically identified velocity offsets, we find that seven of
  the far-infrared luminous galaxies have companions, suggesting that
  they are undergoing interactions/mergers and from their relative
  velocities we can determine a dynamical mass of
  $1.5\pm0.9\times10^{11}M_{\odot}$. These measurements are comparable
  to millimeter CO estimates for the dynamical masses of these systems
  on similar scales, and larger than recent estimates of the dynamical
  masses of UV-selected galaxies at similar redshifts derived in an
  identical manner.  Using the [N{\sc ii}]/H$\alpha$ index to predict
  abundances, we investigate the Luminosity--Metallicity relation for
  these galaxies and find that many have metallicities consistent with
  UV-selected high-redshift galaxies and slightly lower than local
  luminous infrared and elliptical galaxies (although we caution that
  our metallicity estimates have possible systematic uncertainties).
  We also compared our H$\alpha$ and far-infrared luminosities with
  deep {\it Chandra} observations of a subset of our survey fields and
  use these data to further assess their AGN content.  We conclude that
  these high-redshift, far-infrared luminous galaxies represent a
  population of massive, metal-rich, merging systems with high
  instantaneous star formation rates, strong dust obscuration and
  actively-fueled AGN which are likely to be the progenitors of
  massive local elliptical galaxies.
\end{abstract}


\section{Introduction}

Recent surveys in the submillimeter (sub-mm), millimeter (mm) and radio
wave bands suggest that the star formation density detectable by these
dust-independent tracers has evolved strongly with redshift.  Indeed,
this evolution appears to out-strip that found using tracers which are
more sensitive to dust obscuration, suggesting that an increasing
proportion of activity in more distant galaxies may be highly obscured
(Blain et al.\ 1999b).  The populations resolved in these wavebands
appear to be responsible for much of the energy density in the
extragalactic far-infrared/sub-mm background (Smail et al.\ 2002; Cowie
et al.\ 2002; Chapman et al.\ 2004a).  However, the extreme faintness
of optical counterparts to these obscured galaxies has made it very
difficult to obtain accurate redshifts and measure intrinsic properties
(e.g.,\ Simpson et al.\ 2004).

The best-studied examples of the high-redshift, far-infrared luminous
galaxy population are those identified in the sub-mm wave band using
the SCUBA camera on the JCMT.  The median redshift for this population
is $<\!z\!>\sim2.4$ (Chapman et al.\ 2003a, 2003b, 2004a), and their
sub-mm fluxes suggest that they have bolometric luminosities
$>10^{12}L_{\odot}$ -- implying that they are Ultraluminous Infrared
Galaxies (ULIRGs).  The nature of this population and their relevance
to models of galaxy formation models and evolution are particularly
important (e.g.,\ Genzel et al.\ 2003; Baugh et al.\ 2004).  If they
are powered purely by star-formation then these galaxies form about
half of the stars seen locally (Lilly et al.\ 1999).  However, both AGN
activity and star-formation could contribute to their immense
far-infrared luminosities, and without further information it is
impossible to disentangle the precise energy source (Alexander et al.\ 
2003a).

Rest-frame optical emission line properties provide a powerful tool to
investigate star formation rates (SFRs), power sources, and metallicity
of galaxies.  In particular, the Hydrogen Balmer emission line series
is one of the primary diagnostics of the SFR in nearby galaxies, with
the strength of H$\alpha$ and its relative insensitivity to extinction
making it the line of choice.  The nebular recombination lines are a
direct probe of the young, massive stellar population, since only stars
with masses $\gsim$10M$_{\odot}$ and lifetimes $\lsim20$\,Myr
contribute significantly to the integrated ionising flux.  Hence, the
strength of these emission lines provides a nearly instantaneous measure
of the SFR, independent of the previous star formation history.
Moreover, by combining the SFR inferred from this diagnostic line with
the far-infrared emission (which comes from reprocessed radiation which
has been absorbed and re-emitted by dust in the far-infrared at
wavelengths of 10--300$\mu$m), we can gauge the prevalence of dust
obscuration. The width of the H$\alpha$ line and its intensity relative
to other rest-frame optical lines can also give important information
about the presence and luminosity of an AGN within a galaxy (Veilleux
et al.\ 1987, 1995).

H$\alpha$ is visible in the near-infrared wave band out to $z\sim 2.6$
and projects exploiting the H$\alpha$ emission at these high redshifts
have provided unique insights into the star formation properties of
distant galaxies (e.g.,\ Yan et al.\ 1999; Erb et al.\ 2003; Shapley et
al.\ 2004; van Dokkum et al.\ 2004).  With precise redshifts for the
far-infrared luminous population from the work of Chapman et al.\ 
(2004a, 2004b), we can efficiently target the H$\alpha$ emission from
the galaxies to understand the formation and evolution of this
population and can identify the power sources, star formation rates,
metallicities (which are accessible through the [N{\sc ii}]/H$\alpha$
[N2] index; Pettini \& Pagel 2004), and masses, as well as more general
issues such as their relation to other classes of high-redshift sources
such as Lyman Break Galaxies (LBGs; Pettini et al.\ 2001; Erb et al.\ 
2003; Shapley et al.\ 2004).

As well as measuring the star formation rate from H$\alpha$, it is also
possible to measure dynamics of these systems from the same spectra.
Far-infrared luminous galaxies at $z\sim2$ appear morphologically
complex (Chapman et al.\ 2003b; Smail et al.\ 2004). By measuring the
internal dynamics of the galaxies or the velocity offsets between
companions, we can also place limits on their masses and so test if
these galaxies are truly massive systems (Genzel at al.\ 2003).

In this paper we present the results from a near-infrared study of a
sample of far-infrared- detected galaxies at $z$=1.4--2.7.  We use the
H$\alpha$ emission line to derive their star formation rates and
dynamics.  We investigate the metallicity of these galaxies through
their [N{\sc ii}]/H$\alpha$ emission and dynamics as traced by
velocity structures visible in H$\alpha$ emission in a subset of
galaxies.  Using {\it Chandra} data, we also compare our H$\alpha$ and
far-infrared luminosities with X-ray luminosities in several of our
survey fields.

In \S2 we present the data reduction and results from the spectroscopic
survey and narrow-band imaging.  In \S3 we discuss the H$\alpha$
properties of the far-infrared luminous galaxies and present a
discussion of the star formation rates estimated from their H$\alpha$
and far-infrared emission and of their dynamics, metallicities, and
X-ray counterparts.  We present our conclusions in \S4.  We use a
cosmology with $H_{0}=72\kms$, $\Omega_{0}=0.3$ and $\Lambda_{0}=0.7$
in which 1$''$ corresponds to 8.2\,kpc at $z=2.4$.

\section{Observations and Analysis}

Our target sample comes from two catalogs of far-infrared luminous
galaxies by Chapman et al.\ (2003a, 2004a).  The majority of our sample
comprises sub-mm detected, radio-identified galaxies (SMGs).  These
galaxies have precise positions from the microjansky ($\mu$Jy) radio
emission and are confirmed to be far-infrared luminous from their
detection in the sub-mm/mm wavebands with SCUBA or MAMBO.  The
radio-detected subset of the SMG population represents $\sim$60\%--70\%
of all SMGs brighter than $\gs 5$\,mJy (Ivison et al.\ 2002; Chapman et
al.\ 2001; Wang et al.\ 2004).  In addition, we have also included a
small number of optical and sub-mm faint radio galaxies (OFRGs) at
similar redshifts which have been proposed to be similarly luminous
far-infrared galaxies, but with somewhat hotter dust temperatures
resulting in them having comparatively faint submm fluxes (Chapman et
al.\ 2004b; Blain et al.\ 2004).  The median redshift of the combined
sample is $<\!z\!>=2.4\pm0.2$.  For the purposes of this study we chose
targets whose redshifts place H$\alpha$ in spectral regions which are
relatively free from strong atmospheric absorption and emission.

We have explored two routes to investigate the H$\alpha$ emission from
SMGs/OFRGs.  First, we used narrow-band imaging to assess the H$\alpha$
fluxes of galaxies in our sample -- from both a tunable filter (NSFCAM
on IRTF) and more traditional narrow-band filters (with the UFTI
near-infrared imager on UKIRT) for galaxies whose redshifts
serendipitously place H$\alpha$ in the filter bandwidth.  These
narrow-band observations provide the opportunity to determine the total
H$\alpha$ emission from the SMGs/OFRGs and search for any extended
emission or spatial companions.

Second, we used classical long-slit near-infrared spectroscopy with the
NIRSPEC spectrograph on Keck and the ISAAC spectrograph on the VLT.
These observations allow us to measure precise systemic redshifts for
these SMGs/OFRGs, with much higher reliability than that available from
their rest-frame UV emission -- which frequently show velocity shifts
of 100's of km\,s$^{-1}$.  These precise redshifts are necessary for
interferometric CO follow-up of these galaxies (Neri et al.\ 2003;
Greve et al.\ 2004).  The long-slit spectroscopic observations also
allow us to measure the H$\alpha$ luminosities and SFR(H$\alpha$) and
to gauge the prevalence of AGN in these galaxies -- through the
detection of broad lines and extreme [N{\sc ii}]/H$\alpha$ flux ratios.

These two approaches therefore provide complimentary information on the
H$\alpha$ emission properties of the SMG/OFRG population.

\subsection {Narrow-band Imaging}

\subsubsection{IRTF}
Narrow-band imaging of five targets was carried out using the NASA
Infra-Red Telescope Facility\footnotemark (IRTF) 3-m Telescope between
2003 April 28 and May 02.  The observations were made in generally
photometric conditions and $\sim 1''$ seeing.  We used the NSFCAM
camera (Shure et al.\ 1993) which employs a $256\times256$ InSb
detector at 0.15$''$\,pixel$^{-1}$ to give a 38$''$ field of view
(which probes roughly 300\,kpc at $z\sim 2.4$).  The continuously
variable tunable narrow-band filter (CVF) in NSFCAM provides an $R=90$
passband which was tuned to the galaxy redshifts measured from the UV
spectra from Chapman et al.\ (2003a, 2004).  Shorter, matched
broad-band imaging was interspersed between the narrow-band exposures
to provide continuum subtraction.

\footnotetext{The Infrared Telescope Facility is operated by the
  University of Hawaii under Cooperative Agreement no.\ NCC 5-538 with
  the National Aeronautics and Space Administration, Office of Space
  Science, Planetary Astronomy Program.}

The observations were taken in a standard nine-point dither pattern and
reduced using a running flat field of the six nearest temporally
adjacent frames, masking bright objects before creating the flat-field
frame.  The final image was made by averaging the flat-fielded frames
with a $3\sigma$ clip to reject cosmic rays.  To calibrate our data, we
observed UKIRT faint photometric standards (Hawarden et al.\ 2001).
These standards were observed at similar air masses and using the same
instrumental configuration as the target galaxies.  As well as
providing good flux calibrations, these observations also allow us to
calculate the relative throughput of the narrow-band and broad-band
filters.  This allows precise subtraction of the continuum contribution
from the H$\alpha$ emission in the narrow-band filter.  The NSFCAM CVF
filter is wide enough that it contains both the H$\alpha$ and [N{\sc
  ii}] emission lines (with the [N{\sc ii}] emission line included at
95\% of peak transmission assuming that H$\alpha$ falls at the peak of the
filter trace).  The median [N{\sc ii}]/H$\alpha$ ratio in our
spectroscopic sample is 0.25, and we therefore apply a 24\% correction
to the narrow-band flux to account for the [N{\sc ii}] emission line.
Exposure times, central wavelengths, and measured H$\alpha$ fluxes from
these observations are given in Tables~1 \& 2.  We illustrate the
H$\alpha$ morphology for one of the more unusual galaxies in our sample
in Fig.~4 and discuss this in more detail in \S3.1

\subsubsection{UFTI}

We obtained classical narrow-band imaging of two SMGs, SMM\,J105226.61
and SMM\,J131232.31, whose redshifts should place H$\alpha$ emission in
the Br$\gamma$ and H$_2$\,S1 filters, respectively.  These data were
taken with the UFTI near-infrared camera (Roche et al.\ 2003) on
UKIRT\footnotemark on 2003 February 25--27 in $\ls0.7''$ seeing and
photometric conditions. The UFTI camera has a $1024\times1024$ HgCdTe
array and a plate scale of 0.091$''$ pixel$^{-1}$, giving a field of
view of 92$''\times$92$''$ (which probes roughly 0.75 Mpc in our
adopted cosmology).  To provide continuum subtraction, we interspersed
the narrow-band observations with $K$-band imaging.  The data were
reduced using the {\sc orac-dr} pipeline.  The relative throughputs of
the broadband and narrowband filters and flux calibration were
determined using Two Micron All Sky Survey (2MASS) $K$-band photometry
on bright ($K\sim 12$) stars within the 92$''$ field of view.

%
%
\begin{center}
{\footnotesize
{\centerline{\sc Table 1: Summary of Narrow Band Imaging}}
  \smallskip
\begin{tabular}{lcccc}
\hline
\noalign{\smallskip}
~\hspace{1.0cm}Object & $\lambda_{\rm cen}$ & \multispan2{\hfil t$_{\rm exp}$~(ks)\hfil }\\
                      & ($\mu$m)            &                       $K$  &  H$\alpha$  \\
\hline
\hline
\noalign{\smallskip}
~\hspace{-0.3cm}IRTF Tunable Filter\\
SMM\,J105230.73+572209.5          & 2.3692          & 2.5 & 12.6 \\
SMM\,J123635.59+621424.1          & 1.9721          & 1.4 & 9.0  \\
SMM\,J131215.27+423900.93        & 2.3396          & 1.4 & 7.2  \\
SMM\,J140104.96+025223.5          & 2.3396          & 0.7 & 9.0  \\
SMM\,J163631.47+405556.9          & 2.1544          & 4.5 & 14.4 \\
~\hspace{-0.3cm}UKIRT Narrow-band \\
SMM\,J105226.61+572113.0          & 2.248           & 5.6 & 22.7 \\
SMM\,J131232.31+423949.5          & 2.166           & 4.0 & 15.0 \\
\hline
\end{tabular}
\medskip
\begin{minipage}{4.5in} 
  \smallskip Notes: $\lambda_{cen}$ denotes the central position of the
  narrow-band filter for \\
  the observations.  \\
\end{minipage} }
\end{center}

All of our narrow-band observations were taken based on the rest-frame
UV redshifts and assuming that the rest-frame optical redshifts were in
agreement.  However, a velocity offset of $\sim500\kms$ between the
rest-frame UV and optical emission lines can redshift the H$\alpha$ out
of the narrow-band filter.  Shifts of the amplitude are relatively rare,
and so we assume that our observations have sampled the bulk of the
H$\alpha$ emission in these galaxies.

\footnotetext{The United Kingdom Infrared Telescope is operated by the
  Joint Astronomy Centre on behalf of the U.K. Particle Physics and
  Astronomy Research Council.}

\subsection {Keck Spectroscopy}

We obtained near-infrared spectra of 24 SMGs/OFRGs using NIRSPEC
(McLean et al.\ 1998) on the 10-m Keck telescope.\footnotemark These
observations were obtained on the nights of 2003 August 3, in
non-photometric conditions; in somewhat better conditions on 2004
February 1; in $0.5''$ seeing and in photometric condition between 2004
April 6--9 and 2004 July 31.  In all four instances, observations were
made using the low-resolution, longslit mode with a 4-pixel wide
(0.76$''$) slit.  Spectra were obtained in the $K$-band using the N7 or
$K'$ filters.  In this configuration the resolution is $R\sim 1500$. To
subtract sky emission, the observations were made in a standard ABBA
sequence, where the object is nodded along the slit by 10--20$''$.  The
total integration times are listed in Table~2 (the integration times of
individual exposures were typically $\sim 600$s).  One goal of these
observations was to provide precise systemic redshifts of the
SMGs/OFRGs to compare to those measured in the UV and also to provide a
precise redshift for our on-going CO mapping program on the Plateau de
Bure interferometer (Neri et al.\ 2003; Greve et al.\ 2004).  To
maximise the sample size, the exposure times were therefore kept short
and so the H$\alpha$ lines are rarely detected with sufficient
signal-to-noise ratio to measure more than their most basic
characteristics.  We note that the success rate for detecting H$\alpha$
emission in the SMGs/OFRGs in the near-infrared is $\sim70\%$.  This
estimate is based on the results from the third observing run, which
was largely photometric since the first two observing runs were plagued
by clouds.  In cases in which there was no detection of either
continuum or H$\alpha$ emission, we have assumed that the target was
not on the slit (as a result of an offset error or loss of guiding
during the offset from a nearby bright star).  In two cases we detected
continuum from the target but could not detect any emission lines.  In
these cases we have quoted upper limits for the H$\alpha$ luminosity
and SFR.

\footnotetext{Obtained at the W.M. Keck Observatory, which is operated
  as a scientific partnership among the California Institute of
  Technology, the University of California and the National Aeronautics
  and Space Administration. The Observatory was made possible by the
  generous financial support of the W.M. Keck Foundation.}

The spectroscopic observations were reduced using the {\sc wmkonspec}
package in {\sc iraf}.  We remap the two dimensional spectra using
linear interpolation to rectify the spatial and spectral dimensions.
After subtracting pairs of nod-positions, residual sky features were
removed in {\sc idl} using sky regions on either side of the object
spectrum.  The wavelength calibration from the 2003 run used the night
sky lines, while for the 2004 observing runs we used an argon arc
lamp.  The output pixel scale is 4.3\,\AA\,pixel$^{-1}$, and the
instrumental profile has a FWHM of 15\AA\ (measured from the widths of
the sky-lines).  In all line widths quoted in the following sections we
have deconvolved the instrumental profile from the FWHM of the
galaxies.

The redshifts of our targets have been relatively well constrained from
rest-frame UV spectroscopy carried out with LRIS on the Keck telescope.
However, again we stress that the rest-frame UV and optical emission
lines can show velocity offsets of a few 100's\,$\kms$ (Pettini et al.\ 
2001; Erb et al.\ 2003; Shapley et al.\ 2003; Smail et al.\ in prep).
With knowledge of the UV redshift we were able to identify the
H$\alpha$ and [N{\sc ii}] emission in each spectrum and ensure the
correct line identification.

As the first two and final observing runs were non-photometric, we
calibrate the integrated H$\alpha$ fluxes onto an absolute flux scale
by careful comparison of the spectra with the $K$-band photometry of
the galaxies (Smail et al.\ 2004).  We convolve the galaxy spectra with
the normalised $K$-band filter response curve to obtain the mean flux
conversion factor for each galaxy spectrum.  We also note that the
atmospheric transmission at the wavelength of redshifted H$\alpha$ is
$\sim1$ for all targets except SMM\,J123621.27, where the transmission
is $0.75\pm 0.02$.  For the third observing run we used a standard star
(FS 27) to flux-calibrate the data.  We also use the broadband
photometry to flux-calibrate these data in order to verify that the
process applied to calibrate the previous observations is reliable and
obtain a mean ratio of calibration factors of $1.05\pm0.10$.  This
gives us confidence that using the photometry to flux-calibrate gives
consistent results.  All of the objects included here have detectable
continuum emission.  Where the continuum or emission lines are faint,
we binned the reduced two dimensional frames by a factor of two in both
the spatial and spectral dimension to improve the contrast prior to
extraction.  We show the extracted spectra for the whole sample in
Fig.~1.

%
%
\begin{figure*}[tbh]
\centerline{\psfig{file=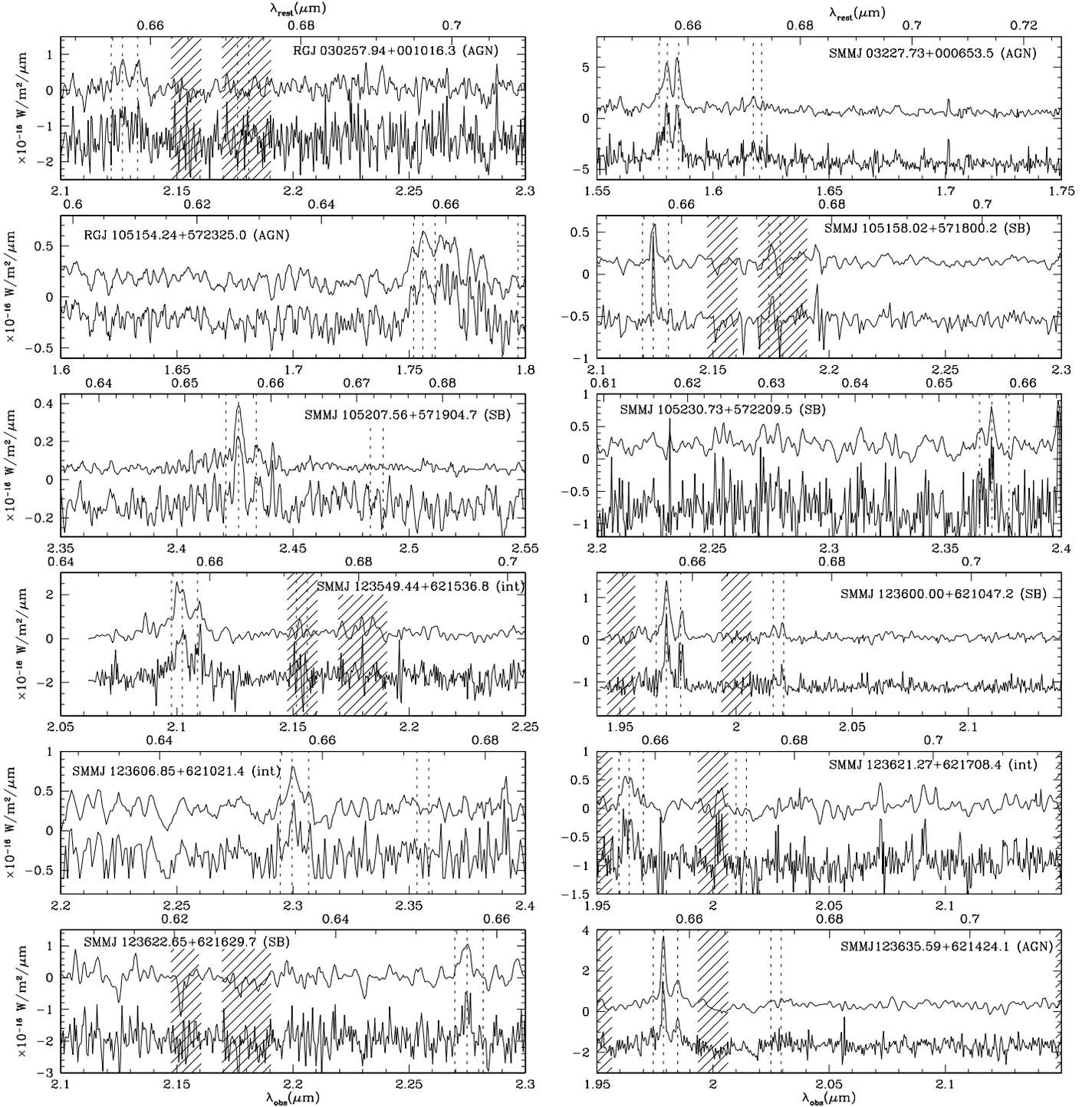,angle=0,width=8.0in}}
\caption{\scriptsize H$\alpha$ spectra for the SMGs/OFRGs in our
  sample.  The lower spectrum in each panel is the raw spectrum (offset
  in flux scale for clarity).  The upper spectrum is smoothed to the
  instrumental resolution.  The dashed vertical lines show the
  positions of the H$\alpha\lambda$6562.8, [N{\sc
    ii}]$\lambda\lambda$6548.1,6583.0, and [S{\sc
    ii}]$\lambda\lambda$6717,6731, emission lines.  The lower axis
  scale shows the observed wavelength, while the upper scale displays
  the rest-frame wavelength.  The shaded areas are regions of strong
  sky emission. The spectra are ordered in right ascension range as in
  Table~2, and we label each galaxy with its spectral classification
  from \S2.4.  Note that we identify two H$\alpha$ emission lines from
  SMMJ123707.21 rather than H$\alpha$ + [N{\sc ii}] (see \S3.1).  The
  final spectra are higher resolution ISAAC/VLT spectra (Table~2). }
\end{figure*}

%
%
\begin{figure*}[tbh]
\centerline{\psfig{file=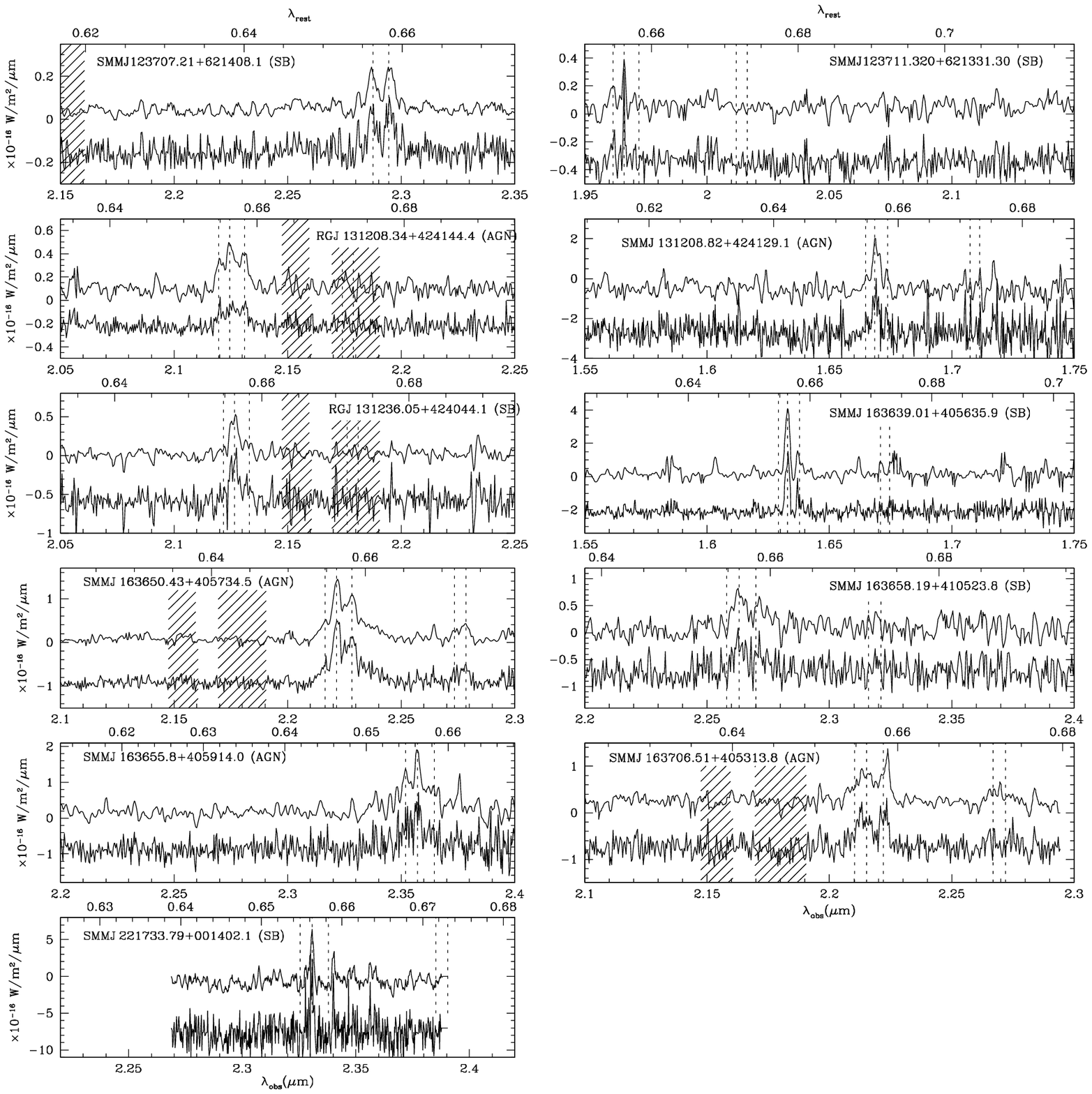,angle=0,width=8.0in}}
\end{figure*}

\subsection {ISAAC VLT Spectroscopy}

During the night of 2003 October 21, observations of SMM\,J221733.79
were obtained in queue mode with the ISAAC spectrograph on the
VLT\footnotemark (Moorwood 1997).  The data were taken using the
medium-resolution ($R=3000$) grating and a 1$''$ slit (with an
instrumental FWHM of $\sim$6.7\AA) and in 0.8$''$ seeing.  In this
configuration the output pixel scale is 1.2\AA\,pixel$^{-1}$.  The total
integration time was 3.0\,ks with the data taken in $4\times750$s
exposures in the standard ABBA sequence where the object is nodded
along the slit by 10--20$''$ to achieve sky subtraction.  We reduced
the data using the ISAAC data-reduction pipeline, which rectifies,
flat-fields, and wavelength calibrates the two dimensional frames.
Prior to extraction, the data were binned by a factor of two in the
spectral direction to boost the contrast of the object.  We observed
the standard star HIP 032007 for flux calibration.  The derived
H$\alpha$ fluxes and redshifts are listed in Table~2, and the spectra
are shown in Fig.~1.

\footnotetext{Based on observations collected with the ESO VLT-UT1 Antu
  Telescope (072.A-0156)}

\subsection{Spectral Analysis}

To accurately determine the redshift for each galaxy, we fit both the
continuum level and emission lines: H$\alpha\lambda$6562.8 and [N{\sc
  ii}]$\lambda\lambda$6548.1,6583.0 simultaneously with a flat
continuum plus Gaussian profiles using a $\chi^{2}$ fit and taking into
account the greater noise in regions of strong sky emission.  As well
as fitting the narrow-line H$\alpha$ and [N{\sc ii}] emission, we also
attempt to identify any underlying broad-line H$\alpha$ (AGN) component
by fitting a broad Gaussian profile at the same redshift as the
narrow-line H$\alpha$ emission but only accepting the result if the
$\chi^2$ fit is significantly better (with $>$90\% confidence limit)
than with no broad-line component.  This allows us to deconvolve the
narrow-line H$\alpha$ flux, which may arise from star-formation, from
the broad-line H$\alpha$ flux from an AGN.  For the observations which
were flux-calibrated using the $K$-band photometry we estimate the
errors in the H$\alpha$ flux and continuum levels using both the
uncertainty in the $K$-band magnitude and the errors on the best fit to
the emission line, evaluated by varying the fit by $\Delta\chi^{2}=1$.
The error bars on the emission-line fluxes and equivalent widths are
therefore conservative errors which take into account the error in
using the $K$-band photometry and the uncertainty which arises from the
signal-to-noise ratio in the data.  This flux error is propagated
through to the star formation rate.  We list the parameters for the
fits for narrow (and where relevant broad) components in Table~2, along
with their uncertainties.  We note that the [S{\sc
  ii}]$\lambda\lambda$6716,6731 doublet is detected in at least three
individual spectra: SMM\,J123600.00, SMM\,J163650.43, and
SMM\,J163639.01.

The ratio of [N{\sc ii}] to H$\alpha$ line fluxes from our spectra, as
well as the presence of a broad component, can be used to identify
luminous AGN in these galaxies (e.g.,\ Veilleux et al.\ 1987; Armus et
al.\ 1989).  We measure all of these observables from our spectra and
use them to flag galaxies whose H$\alpha$ and far-infrared luminosities
could be affected by emission from an AGN, rather than being
star-formation dominated (Table~2).  Using the [N{\sc ii}]/H$\alpha$
emission line ratio and H$\alpha$ line width as diagnostics, we
separate our sample into three classes: AGN dominated, intermediate,
and systems whose properties are consistent with star-formation (but
may still contain some AGN).  In a later section we test the
reliability of these classes through the use of deep X-ray
observations.  We classify galaxies whose near-infrared spectra show
[N{\sc ii}]/H$\alpha$$<$0.7 and FWHM$_{\rm rest}$$<$500km\,s$^{-1}$ as
star-bursts (SB).  Galaxies with FWHM$_{\rm rest}$ between 500 and
1000km\,s$^{-1}$ are classified as intermediate, while galaxies with
[N{\sc ii}]/H$\alpha$$>$0.7 and/or FWHM$_{\rm rest}$$>$1000km\,s$^{-1}$
are classified as AGN.  Using this classification, $\sim$40\% (9/24) of
the objects in our detected spectroscopic sample have some indication
of an AGN (most classifications are based on H$\alpha$ line widths,
with only one AGN changing classification if we remove the limits on
[N{\sc ii}]/H$\alpha$).  This is similar to the rate estimated from UV
spectroscopy and high-resolution radio imaging (Chapman et al.\ 2003a,
2003b, 2004c).  By comparing the rest-frame UV and optical spectral
classifications (Table~2), we find that the fifteen galaxies identified
as SB in the rest-frame UV are classified as SB[9], AGN[3] and
intermediate[3] from their rest-frame optical spectra.  Likewise, the
three AGN identified in the rest-frame UV are all classified as AGN
from their rest-frame optical spectra. We conclude that these
classifications are in reasonable agreement.  However, we note that
their AGN could be easily hidden from view in both wavebands (see
e.g.,\ \S3.5).  Looking at the far-infrared luminosities of the three
spectral classes we find median values of $5.2\pm 3.0\times
10^{12}$\,L$_\odot$ for the SMGs/OFRGs classed as AGN, $6.6\pm
0.7\times 10^{12}$\,L$_\odot$ and $5.4\pm 1.4\times 10^{12}$\,L$_\odot$
for the starbursts and intermediate respectively. We thus find no
strong evidence for a strong luminosity dependence of the different
spectral classes.

To determine total H$\alpha$ fluxes, we have also corrected for slit
losses based on the average $K$-band light distribution of the
galaxies.  As we know the position, width, and orientation of the
spectroscopic aperture for each galaxy, we use the $K$-band images to
calculate the fraction of the total $K$-band light that enters the
spectroscopic aperture.  The galaxies are frequently extended and
disturbed (e.g.,\ Smail et al.\ 2004), and while this correction is
uncertain, we view it as more reliable to apply this factor before
comparing H$\alpha$ to far-infrared star formation rates, rather than
to ignore it.  However, we note that using a single factor for
correcting slit losses may contribute some scatter when comparing the
H$\alpha$ and far-infrared star formation rates.  By careful comparison
of a simulated slit for each galaxy in our sample with the $K$-band
imaging, we estimate the fraction of the flux entering the slit
compared to the total $K$-band flux of the galaxy, $f$, and obtain a
mean value of $f=0.62\pm 0.06$.  We assume that the equivalent width of
the H$\alpha$ emission line is constant across the galaxy and so only
those galaxies with spectra taken in photometric conditions, whose flux
calibration was carried out using the standard star observations,
require this correction to their H$\alpha$ flux.

Finally, we note that there are two repeated observations of targets
between the spectroscopic and narrow-band imaging samples:
SMM\,J105230.60 and SMM\,J123635.59 --- as well as narrow-band imaging
of SMM\,J140104.96 for which an H$\alpha$ spectrum exists in the
literature ($0.56\pm 0.08\times 10^{-18}$\,W\,m$^{-2}$; Ivison et al.\ 
2000).  The agreement between the H$\alpha$ fluxes from the two
techniques is very good (Table~2), with a median ratio of spectroscopic
to imaging fluxes of $0.96\pm 0.04$.  On the basis of this good
agreement we feel confident in combining the spectroscopic and
narrow-band imaging data to discuss the H$\alpha$ luminosities of
SMGs/OFRGs in the following sections.

%
%
\begin{table*}
{\footnotesize
\begin{center}
{\centerline{\sc Table 2: Summary of Results}}
\smallskip
\hspace*{-0.7cm}
\begin{tabular}{lccccccrcc}
\hline
\noalign{\smallskip}
~\hspace{1.0cm}Object & t$_{\rm exp}$ &  $z$ & H${\alpha}$~Flux & FWHM$_{\rm rest}$ & EW$_{\rm rest}$ & SFR(H$\alpha$) & [N{\sc ii}]/H$\alpha$ & L$_{\rm FIR}$ & Class \\

                      & (ks)          &   &(10$^{-19}$\,W\,m$^{-2}$) & (km\,s$^{-1}$) & (\AA)           & (M$_{\odot}$\,yr$^{-1}$) &   & (10$^{12}$L/L$_{\odot}$) & (UV / H$\alpha$) \\
\hline
\hline
\noalign{\smallskip}
~\hspace{-0.3cm} {\bf Spectroscopy}\\
~\hspace{-0.3cm} NIRSPEC Keck \\

SMM\,J030227.73+000653.5  & 3.0  & $1.4076[2]$  & $15.2\pm2.0$  & $420\pm15$    & $85\pm10$     & $140\pm18$  & $ 1.38\pm0.07 $   & $ 5.78\,_{-0.82}^{+2.44}  $     &  SB / AGN     \\

RGJ\,030258.94+001016.3   & 2.0  & $2.2404[8]$  & $1.8\pm0.5$   & $327\pm22$    & $360\pm110$   & $51\pm15$   & $1.13\pm0.40$     & $ 7.74\,_{-1.41}^{+1.41}  $    & int / AGN      \\

RGJ\,105154.24+572325.0   & 4.8  & $1.681[8]$   & $2.7\pm1.9$   & $1565\pm250$  & $91\pm6$      & $61\pm40$   & $ <0.2 $          & $ 2.80\,_{-0.35}^{+0.35}  $   & SB / AGN        \\

SMM\,J105158.02+571800.3  & 2.4  & $2.2390[4]$  & $2.4\pm1.2$   & $257\pm44$    & $21\pm5$      & $57\pm15$   & $ <0.1 $          & $ 10.40\,_{-1.00}^{+1.90} $   & SB / SB         \\

SMM\,J105207.56+571904.7  & 4.8  & $2.692[2]$   & $1.3\pm0.4$   & $285\pm20$    & $21\pm4$      & $217\pm64$  & $ 0.18\pm0.10 $   & $ 9.46\,_{-2.70}^{+2.70}  $   & SB / SB         \\

SMM\,J105230.73+572209.5  & 2.4  & $2.6100[3]$  & $1.2\pm0.3$   & $171\pm40$    & $10\pm3$      & $42\pm15$   & $ <0.05 $         & $ 10.29\,_{-1.12}^{+0.75} $   & SB /  SB        \\

SMM\,J123549.44+621536.8  & 2.4  & $2.2032[3]$  & $15\pm1$      & $536\pm33$    & $184\pm9$     & $239\pm18$  & $0.50\pm0.10$     & $ 6.76\,_{-1.12}^{+1.50}  $   & SB / int        \\

SMM\,J123600.15+621047.2  & 1.2  & $2.0017[2]$  & $3.7\pm0.3$   & $305\pm12$    & $91\pm8$      & $126\pm8$   & $ 0.20\pm0.10 $   & $ 10.50\,_{-1.50}^{+1.50} $   & SB / SB         \\

SMM\,J123606.85+621021.4  & 2.4  & $2.5054[8]$  & $2.0\pm0.3$   & $612\pm35$    & $28\pm5$      & $78\pm12$   & $ <0.2 $          & $ 8.70\,_{-1.20}^{+1.80}  $   & SB / int        \\

SMM\,J123621.27+621708.4  & 2.4  & $1.9924[7]$  & $2.0\pm0.6$   & $586\pm92$    & $212\pm25$    & $56\pm16$   & $0.20\pm0.15$     & $ 12.76\,_{-2.25}^{+1.50} $    & SB / int       \\

SMM\,J123622.65+621629.7  & 2.4  & $2.4662[5]$  & $3.4\pm0.6$   & $434\pm25$    & $137\pm40$    & $125\pm20$  & $ <0.05 $         & $ 9.01\,_{-2.03}^{+1.88}  $   & SB / SB         \\

SMM\,J123635.59+621424.1  & 1.8  & $2.0150[2]$  & $4.2\pm0.3$   & $240\pm33$    & $45\pm10$     & $130\pm30$  & $0.67\pm0.27$     & $ 7.47\,_{-1.50}^{+1.50}  $   & AGN / AGN       \\
                          &      &              & $6.9\pm0.9$   & $1623\pm213$  & $73\pm10$     &             &                   &                               & Broad line cmpt \\

SMM\,J123707.21+621408.1  & 3.6  & $2.490[5]$   & $1.6\pm0.5$   & $348\pm40$    & $35\pm8$      & $88\pm24$   & $ ... $           & $ 5.48\,_{-1.58}^{+2.03}  $   & SB /SB          \\

SMM\,J123711.32+621331.3  & 2.4  & $1.9958[4]$  & $0.4\pm0.3$   & $112\pm18$    & $10\pm5$      & $16\pm9$    & $ <0.05 $         & $ 5.48\,_{-1.58}^{+2.03}  $   & int / SB        \\ 

SMM\,J123711.98+621325.7  & 2.4  & $...$        & $<0.25$       & $...$         & $<5$          & $<6$        & $... $            & $ 3.88\,_{-0.82}^{+1.05}  $   & SB / ...        \\ 

SMM\,J131205.60+423946.0  & 2.4  & $...$        & $<0.3$        & $...$         & $<6$          & $<8$        & $...$             & $ 5.50\,_{-2.00}^{+2.00}  $   & SB / ...         \\

RGJ\,131208.34+424144.4   & 2.4  & $2.2372[18]$ & $0.9\pm0.2$   & $448\pm60$    & $24\pm5$      & $32\pm18$   & $ 1.20\pm0.60 $   & $ 5.13\,_{-1.05}^{+1.05}  $   & SB / AGN        \\
                          &      &              & $0.9\pm0.8$   & $959\pm100$   & $26\pm5$      &             &                   &                               & Broad line cmpt \\

SMM\,J131208.82+424129.1  & 2.4  & $1.5439[6]$  & $6.1\pm2.3$   & $387\pm60$    & $103\pm15$    & $111\pm32$  & $ 0.20\pm0.15 $   & $ 3.23\,_{-0.19}^{+0.15}  $   & SB / AGN        \\
                          &      &              & $2.1\pm1.0$   & $450\pm80$    & $120\pm20$    &             &                   &                               & Broad line cmpt \\

RGJ\,131236.05+424044.1   & 2.4  & $2.2402[8]$  & $2.3\pm1.2$   & $447\pm75$    & $20\pm6$      & $106\pm40$  & $ 0.22\pm0.08 $   & $ 6.69\,_{-0.59}^{+0.59}  $   & SB / SB        \\

SMM\,J163639.01+405635.9  & 2.4  & $1.4880[6]$  & $8.9\pm3.0$   & $248\pm25$    & $188\pm35$    & $147\pm48$  & $ 0.30\pm0.20 $   & $ 5.47\,_{-1.50}^{+1.88}  $   & SB / SB         \\

SMM\,J163650.43+405734.5  & 4.8  & $2.3850[5]$  & $12.2\pm2.0$   & $306\pm47$    & $105\pm20$    & $58\pm19$   & $ 0.41\pm0.10 $   & $ 50.5\,_{-15.8}^{+15.0} $   & int / AGN       \\
                          &      &              & $12.0\pm1.0$  & $1753\pm238$  & $1236\pm200$  &             &                   &                               & Broad line cmpt \\

SMM\,J163655.80+405914.0  & 2.4  & $2.5918[6]$  & $2.4\pm0.4$   & $225\pm29$    & $336\pm60$    & $102\pm15$  & $0.45\pm0.10$     & $ 10.89\,_{-3.75}^{+2.21} $   & AGN / AGN       \\
                          &      &              & $16\pm2$      & $2962\pm402$  & $254\pm40$    &             &                   &                               & Broad line cmpt \\

SMM\,J163658.19+410523.8  & 2.4  & $2.4482[6]$  & $1.9\pm0.4$   & $364\pm77$    & $76\pm15$     & $71\pm35$   & $0.65\pm0.3$      & $ 10.90\,_{-3.77}^{+2.25} $   & SB / SB         \\

SMM\,J163706.51+405313.8  & 2.4  & $2.3745[9]$  & $4.7\pm0.6$   & $225\pm29$    & $96\pm15$     & $160\pm20$  & $0.27\pm0.04$     & $7.17\,_{-3.00}^{+4.21}   $   & AGN  / AGN      \\
                          &      &              & $2.7\pm0.8$   & $3317\pm987$  & $125\pm15$    &             &                   &                               & Broad line cmpt \\

\noalign{\smallskip}
~\hspace{-0.3cm}ISAAC VLT \\

SMM\,J221733.91+001352.1  & 3.0  & $2.5510[7]$  & $8.5\pm3.5$   & $198\pm98$    & $30\pm10$     & $254\pm128$ & $ ... $           & $4.94\,_{-2.63}^{+1.88}   $   &  SB / SB        \\

\noalign{\smallskip}
~\hspace{-0.3cm}{\bf Narrow-band imaging} \\
~\hspace{-0.3cm}NSFCAM IRTF \\

SMM\,J105230.73+572209.5    & 12.6  & $2.610[5]$  & $1.2\pm0.5$   & ... & ... & $21\pm8$   & ... & $10.3\,_{-1.13}^{+0.75}  $ & SB / ...  \\

SMM\,J123635.59+621424.1    & 9.0   & $2.005[5]$  & $8.1\pm1.0$   & ... & ... & $181\pm20$ & ... & $7.47\,_{-1.50}^{+1.50}  $ & AGN / ... \\

SMM\,J131215.27+423900.9    & 7.2   & $2.565[5]$  & $11.8\pm1.0$  & ... & ... & $493\pm41$ & ... & $13.90\,_{-2.48}^{+0.53} $ & AGN / ... \\

SMM\,J140104.96+025223.5    & 9.0   & $2.565[5]$  & $1.2\pm0.2$   & ... & ... & $47\pm10$  & ... & $6.76\,_{-2.50}^{+2.50}  $ & AGN / ... \\

SMM\,J163631.47+405546.9    & 14.4  & $2.283[5]$  & $0.30\pm0.19$ & ... & ... & $10\pm6$   & ... & $10.97\,_{-4.73}^{+6.76} $ & AGN / ... \\

\noalign{\smallskip}
~\hspace{-0.3cm}UFTI UKIRT \\

SMM\,J105226.61+572113.0    & 22.7  & $2.425[5]$  & $0.78\pm0.1$  & ... & ... & $27\pm4$   & ... & $2.40\,_{-1.40}^{+1.40}  $  & SB / ...  \\

SMM\,J131232.31+423949.5    & 15.0  & $2.300[5]$  & $1.9\pm0.1$   & ... & ... & $58\pm5$   & ... & $14.80\,_{-2.93}^{+0.98} $  & SB / ...  \\

\hline

\end{tabular}
\end{center}

\noindent 
Note the value given in the [] $z$ column is the error in
the last decimal place. The H$\alpha$ flux given in column 4 is the
narrow line H$\alpha$ flux unless otherwise stated.  The H$\alpha$ 
fluxes are the observed flux (not corrected for slit flux losses or 
atmospheric transmission), but the SFR's have had both corrections 
applied (where applicable).  The H$\alpha$ star-formation rates are 
uncorrected for extinction. H$\alpha$ classifications are described 
in \S2.4 (SB=star burst; int=intermediate, AGN=AGN).  }
\end{table*}

%
%
\begin{figure*}[tbh]
\centerline{\psfig{file=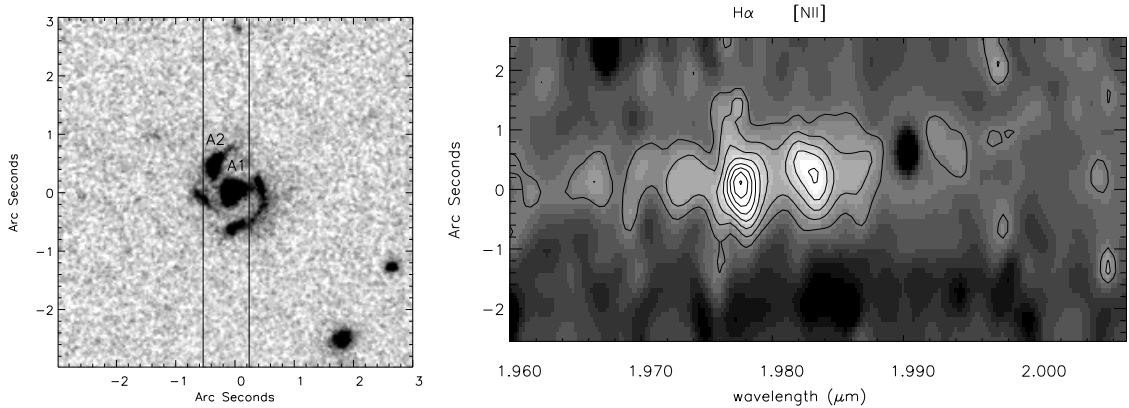,angle=0,width=6.0in}}
\caption{\small {\it Left:} Ccombined $B$, $V$ and
  $I$-band {\it HST} ACS observations of SMM\,J123635.59+621424.1 with
  the slit position overlaid.  {\it Right:} The position-velocity
  diagram around the H$\alpha$ line from the NIRSPEC near-infrared
  spectrum of the galaxy.  This shows an apparent
  $\sim$100\,km\,s$^{-1}$ velocity gradient in the H$\alpha$ emission
  line and a somewhat larger velocity gradient,
  $\sim$150\,km\,s$^{-1}$, in [N{\sc ii}].  The central component of
  the galaxy (labeled A1) has an [N{\sc ii}]/H$\alpha$ ratio of
  $0.60\pm0.15$ and hosts a bright, unresolved radio source (Chapman et
  al.\ 2004a); both features suggest that it contains an AGN.  A1 is
  offset from A2 by $\sim$150\,km\,s$^{-1}$ in velocity and around
  3\,kpc in projection on the sky.  A2 has an [N{\sc ii}]/H$\alpha$
  ratio of $0.2\pm0.1$, suggesting that this is a star-forming knot or
  close companion.  Here 1$''$ corresponds to 8.4\,kpc at the redshift
  of this galaxy and the image has been rotated to match the position
  angle of the spectrum.}
\end{figure*}

\subsection {Star Formation Rates}

For solar abundances and adopting a Salpeter initial mass function
(IMF), the conversion between H$\alpha$ flux and SFR is SFR(M$_{\rm
  \odot}$yr$^{-1}) = 7.9\times10^{-35}$L(H$\alpha$)\,W (Kennicutt et
al.\ 1998).  This calibration assumes that all of the ionising photons
are reprocessed into nebular lines (i.e.,\ they are neither absorbed by
dust before they can ionise the gas, nor do they escape the galaxy).

We also have a second SFR indicator for our sample of SMGs/OFRGs --
their far-infrared (FIR) luminosities. A significant fraction of the
bolometric luminosity of the most active dusty galaxies is absorbed by
interstellar dust and re-emitted in the thermal IR, at wavelengths
10--300$\mu$m.  If young stars dominate the radiation output in the
UV-visible wave band and the dust opacity is high everywhere, then the
far-infrared luminosity measures the bolometric luminosity of the
star-burst and this in turn provides an excellent tracer of the SFR of
the galaxy.  Adopting the models of Leitherer et al.\ (1995) for
continuous bursts of age 10--100\,Myr and using the same IMF as in the
H$\alpha$ calculation yields
SFR(M$_{\odot}$\,yr$^{-1})=4.5\times10^{-37}$L(FIR)\,W (Kennicutt et
al.\ 1998).  We stress that this relation only holds if the age of the
star-burst is less than 100\,Myr.

We list the SFR and H$\alpha$ flux measurements from our sample in
Table~2.  We also give the far-infrared luminosities of these galaxies
from Chapman et al.\ (2003a, 2004b).  These values are derived from
fitting model spectral energy distributions (SEDs) to the observed
850\,$\mu$m and 1.4-GHz fluxes of the galaxies at their known
redshifts, assuming that the local far-infrared--radio correlation
holds (Condon et al.\ 1991; Garrett 2002).

\subsection{Notes on Individual Galaxies}

Observations of six of the galaxies in our sample are particularly
noteworthy and so we discuss these in more detail here. \smallskip

%
%
\begin{figure*}[tbh]
\centerline{\psfig{file=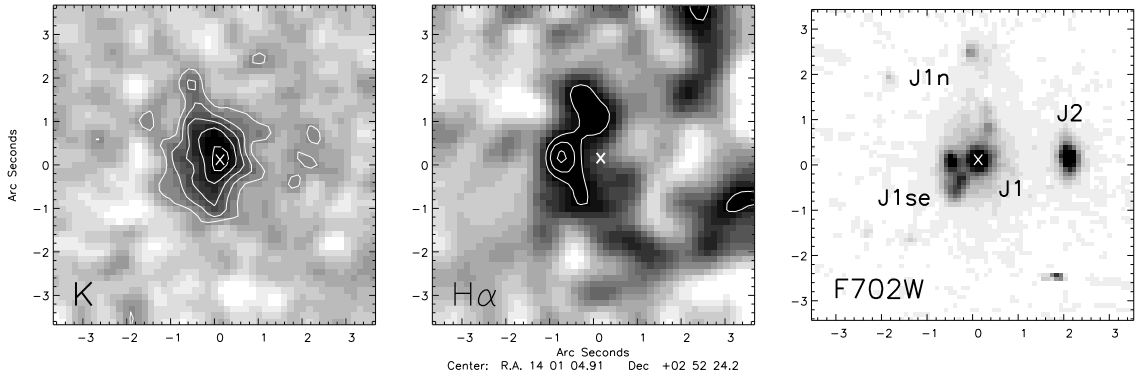,angle=0,width=6.0in}}
\caption{\small Three views of SMM\,J140104.9: broadband $K$ image 
  from IRTF ({\it left}); H$\alpha$ continuum-subtracted, narrow-band
  image from IRTF ({\it middle}) and the {\it HST} $R_{702}$-band of
  the system which illustrates its very complex morphology ({\it
    right}).  We identify in the {\it HST} image the three main
  components of this galaxy as defined by Ivison et al.\ (2001), and in
  each panel we also mark the position of the peak of component J1.  It
  is clear that the optical-near-infrared colors and H$\alpha$ surface
  brightness of component J2 are much different from those of J1n and
  J1se -- ruling out the suggestion that J2 represents a lensed
  counter-image of J1n or J1se (Downes \& Solomon 2003).  Moreover, the
  H$\alpha$ emission extending south from J1n and wrapping around J1
  traces the morphology of the clumps visible in the $R$-band, J1se,
  and has no significant contribution from the smooth component, J1.
  We discuss the interpretation of this result in more detail in \S3.1}
\end{figure*}

\noindent{\it SMM\,J123635.59+621424.1 }--- Dawson et al.\ (2003) identified 
this object as a z=2.015 spiral galaxy, but their near-infrared
spectroscopy around the H$\alpha$ emission line shows a
2500$\pm$250\,$\kms$ broad-line H$\alpha$ component and [N{\sc
  ii}]/H$\alpha$ emission line flux ratio $0.45\pm0.1$.  The presence
of the broad-line component and detection of hard ($\Gamma=0.3$) X-ray
emission from {\it Chandra} imaging indicate an obscured type II AGN
(Dawson et al.\ 2003).  Our spectroscopic and narrow-band observations
of this galaxy produce comparable H$\alpha$ fluxes, which, along with
the compact morphology of the galaxy in the IRTF narrow-band image,
suggests that the spectroscopic slit is sampling the bulk of the
emission from this system (the slightly lower flux in the narrow-band
observation arises because the narrow-filter only samples the
narrow-line H$\alpha$; $\sim20$\% of the broad-line flux is missed).
Our spectroscopic observations of this target show clear velocity
structure in the H$\alpha$ and [N{\sc ii}] emission lines (Fig.~2).  We
aligned the slit with the center of the galaxy and the bright knot
(labeled A2 in Fig.~2), and our resulting Keck spectroscopic
observations indicate either a velocity shear or rotation in the
H$\alpha$ emission line with an amplitude of $\sim100$\,km\,s$^{-1}$
within $\sim2''$ (17\,kpc in projection) along the slit (Fig.~2).  By
collapsing down the H$\alpha$ and [N{\sc ii}] emission lines in the
spectral direction, we find an offset of $0.3''$ (3\,kpc) along the
slit between the maximum intensities of these two lines, as well as an
associated variation in the [N{\sc ii}]/H$\alpha$ ratio from
$\sim0.2\pm0.1$ up to $\sim0.60\pm0.15$ (which is consistent with the
results of Dawson et al.\ 2003).  We also find a similar broad-line
component to the H$\alpha$ emission, confirming the presence of an AGN.
The high-resolution {\it Hubble Space Telescope\footnotemark} ({\it
  HST}) ACS image of this galaxy (Fig.~2), from the GOODS imaging of
this region (Giavalisco et al.\ 2004; Dawson et al.\ 2003; Smail et
al.\ 2004), shows an apparently face-on spiral galaxy with a bright
nucleus (A1) and a prominent companion or knot in one of the spiral
arms (A2); the optical extent of the galaxy is 13\,kpc (1.5$''$).  The
separation between the nucleus (A1) and knot (A2) in the image is
$\sim0.5''$ (5\,kpc) -- comparable to the apparent offset in the
emission-line peaks.  The high-resolution 1.4-GHz MERLIN map of this
galaxy in Chapman et al.\ (2004c) shows an unresolved radio source
coincident with the nuclear component, which is also where the [N{\sc
  ii}]/H$\alpha$ is the strongest ($\sim0.6$), providing further
support for the classification of this component as an AGN.  The
face-on aspect of this system, combined with the modest velocity
difference between A1 and A2, suggests to us that the latter may be a
dynamically-distinct component (rather than a star-forming knot inside
a spiral arm), an interaction which has prompted the activity we see.
We note that it is possible for intensity gradients between separate
components to mimic velocity gradients as a result of the way in which
long-slit spectroscopy mixes spatial and spectral domains.  However,
the spatial offset between A1 and A2 in the dispersion direction is
$\sim 0.4''$, corresponding to 10\AA, which is much less than the
apparent velocity gradient.  We therefore suggest that the velocity
offset most likely arises from motions within the galaxy, rather than
as an artifact of the observation.  \medskip

\footnotetext{Based on observations made with the NASA/ESA {\it Hubble
    Space Telescope} which is operated by STSCI for the Assosciate of
  Universities for Research in Astronomy, Inc., under NASA contract
  NAS5-26555}

\noindent{\it SMM\,J123707.21+621408.1 }--- We observed this
target with a position angle such that the NIRSPEC slit passes through
the two components shown in the {\it HST} ACS imaging in Smail et al.\ 
(2004).  This galaxy consists of a red component and a much bluer
object separated by 0.2$''$ (1.7\,kpc; labeled B1 and B2 in Fig~5).
The resulting NIRSPEC spectrum shows two strong, spatially-extended
lines separated by $\sim600$\,km\,s$^{-1}$.  We identify both of these
emission lines as H$\alpha$ (rather than H$\alpha$ and [N{\sc ii}])
from two separate components for two reasons: (1) the spatial
separation between the two components in the spectra is matched almost
exactly by the separation in the {\it HST} ACS image (Fig.~5); and (2)
if the higher velocity component was identified as strong [N{\sc
  ii}]$\lambda6583$ then the line ratio would be [N{\sc ii}]/H$\alpha$
$\sim1$, and yet we see no signs of the [S{\sc ii}] emission line, which
has a ratio of [S{\sc ii}]/[N{\sc ii}]$\sim0.5$ for AGN (Ferland \&
Ostenbrock 1986).

%
%
\addtocounter{figure}{1}
\begin{figure*}[tbh]
\centerline{\psfig{file=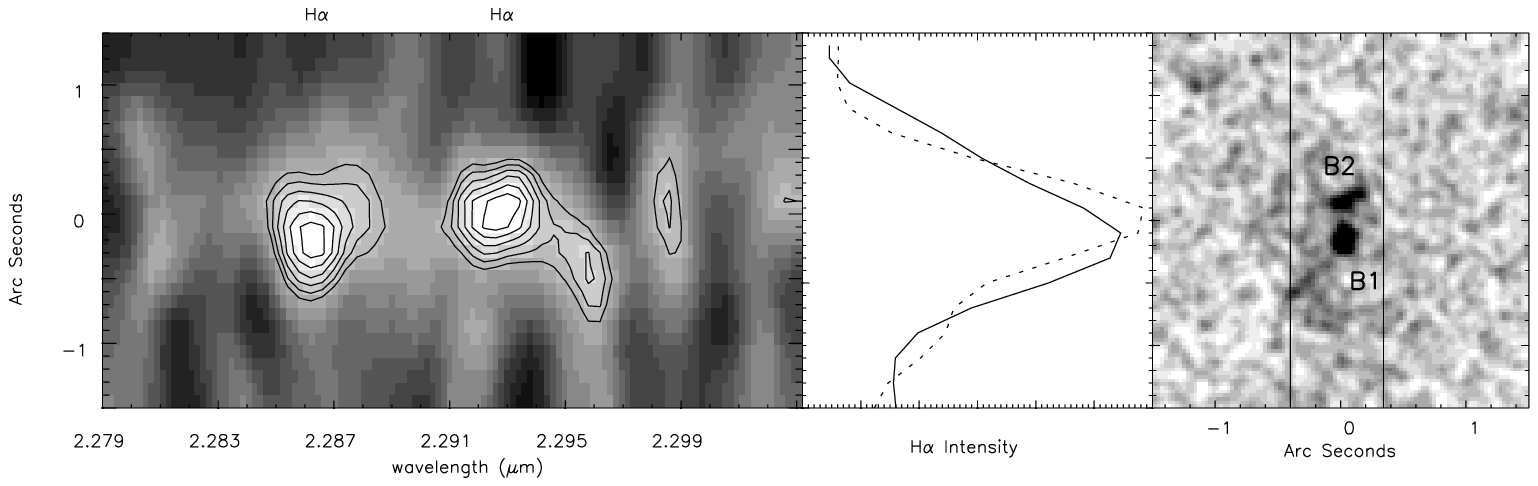,angle=0,width=6.0in}}
\caption {\small {\it Left:} Position-velocity diagram around
  the H$\alpha$ line from the near-infrared spectrum of
  SMM\,J123707.21. {\it Right:} Combined $B-$, $V-$ and $I$-band {\it
    HST} ACS observations of this galaxy with the slit position
  overlaid.  {\it Middle:} Intensity distribution of the H$\alpha$
  emission collapsed over the central $200\kms$ of each of the two
  H$\alpha$ emission lines.  This shows an apparent spatial offset of
  0.2$''$ (and 600\,km\,s$^{-1}$ velocity offset) between the two
  galaxies.  This matches the apparent separation of the central
  component of the galaxy (labeled B1) from the much redder second
  component (labeled B2) which are offset by 2\,kpc in projection.
  (For a color version of this image, see Smail et al.\ 2004).}
\end{figure*}

\medskip

\noindent{\it SMM\,J123711.32+621331.3} --- This target was known to consist of 
two radio sources, both of which may contribute to the far-infrared
emission.  Aligning the slit along both components, we detect only
faint continuum and no lines from the UV-identified component,
SMM\,J123711.98+621325.7, for which Chapman et al.\ (2004a) measured
the redshift of the system.  However, we do detect H$\alpha$ emission
from the second radio source, SMM\,J123711.32+621331.3.  This allowed
us to derive a redshift of $1.9958\pm0.0004$, giving an offset of
$400\pm50\kms$ and 8$''$ ($\sim$70\,kpc) in projection from
SMM\,J123711.98+621325.7.

\medskip

\noindent{\it RGJ\,131236.05+424044.1} --- The ground-based $K$-band imaging 
of this galaxy from Smail et al.\ (2004) shows a bright nucleus
surrounded by a diffuse halo approximately 2$''$ (16\,kpc) in extent.
The two components seen in the near-infrared spectrum are separated by
$\sim185\pm45\kms$ in velocity and 0.4$''$ (3.4\,kpc) in projection.
The two-dimensional near-infrared spectrum suggests that one of the
sources (labeled C1) displays signs of a velocity shear in H$\alpha$
across $\sim0.5''$.  The [N{\sc ii}]/H$\alpha$ ratio shows marginal
evidence for variation from $0.45\pm0.10$ to $0.35\pm0.10$ between C1
and C2 and also mimics the velocity offset between C1 and C2.  These
modest [N{\sc ii}]/H$\alpha$ emission line ratios indicate that both
components are likely to be star-burst, rather than AGN-, dominated.

%
%
\centerline{ \psfig{file=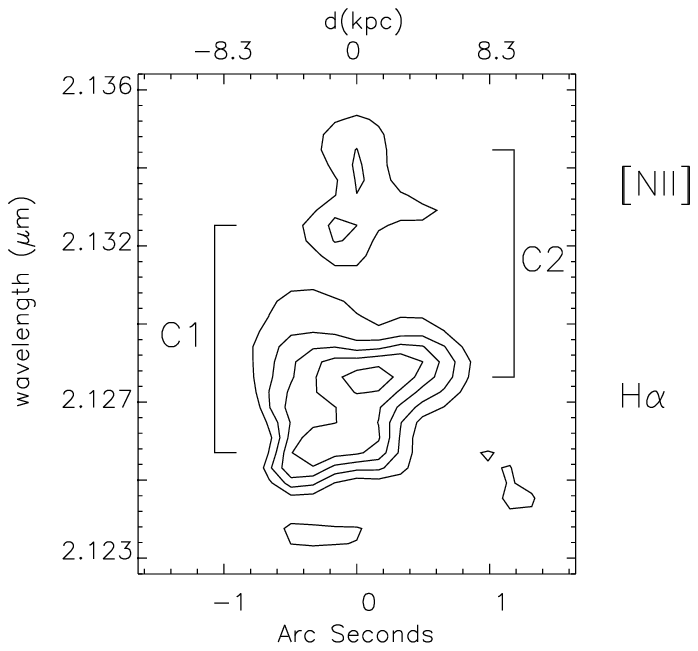,width=3.0in,angle=0}} {\noindent
  \footnotesize\addtolength{\baselineskip}{-2pt} {\sc Fig.~4. --} Two
  Dimensional near-infrared spectrum of RGJ\,J131236.05, showing the
  velocity structure seen in H$\alpha$ and [N{\sc ii}].  We identify
  two components (labeled C1 and C2), offset by
  $185\pm40$\,km\,s$^{-1}$ and 3.4\,kpc in projection along the slit.}

\medskip

\noindent{\it SMM\,J140104.96+025223.5} --- This $z=2.56$ sub-mm selected
galaxy (SMM\,J14011+0252) is discussed in detail by Ivison et al.\ 
(2000, 2001).  It lies in the field of the $z=0.25$ cluster A\,1835 and
is expected to be amplified by a factor of 2.75 by the foreground
cluster potential. The morphology of this galaxy is complex (Fig.~3).
Ivison et al.\ (2001) identify three main components: J1, a blue
relatively smooth and regular object; J2, a bluer and more compact
object about 2$''$ to the west of J1; and J1n, an extremely red, diffuse
structure which extends 2--3$''$ to the north of J1.

The interpretation of this multi-component system is contentious.
Ivison et al.\ (2001) suggest that the gas reservoir and source of the
far-infrared emission from this galaxy reside in J1n -- with J2 and J1
simply being less-obscured components within the same system.  However,
Downes \& Solomon (2003) suggest that the clumpy features to the
south-west of J1 (called J1se), as well as J1n and J2, are all images of
a single, highly-amplified background sub-mm source -- which is being
lensed by both the regular component of J1 (which they identify as a
foreground ($z\sim 0.20$) galaxy) and the cluster potential.

Our IRTF continuum-subtracted H$\alpha$ image provides a powerful tool
for testing these competing suggestions (see also Tecza et al.\ 2004).
We compare the morphology of the H$\alpha$ emission with the $R_{702}$-
and $K$-band images in Fig.~3.  It is clear that the bulk of the
emission traces components J1 and J1se, with the regular J1 and compact
J2 being undetected.  This immediately rules out the possibility that
J2 is a lensed counter-image of J1n or J1se.  However, the absence of
H$\alpha$ emission from the smooth component of J1 is a concern --
suggesting that the claim by Downes \& Solomon (2003) that it is a
foreground lens may be correct.  To further test this, we have returned
to the optical spectrum of J1 presented by Barger et al.\ (1999) and
claimed by them to represent a classical Lyman-break galaxy.  More
careful study of this spectrum leads us to re-examine this
interpretation: there are strong absorption features at 4745, 4784,
4912, 4957, 5374, and 6070\AA, which are unidentifiable at the redshift
claimed by Barger et al., $z=2.55$, but correspond exactly to Balmer
H$\zeta$, H$\eta$, Ca H\&K, G-band and H$\beta$ absorption at
$z=0.248$.  This is an unfortunate conjunction -- J1 is a $\sim
0.1$L$^\ast$, post-starburst member of the A\,1835 cluster.  The
absence of any strong emission lines in this galaxy and broadband
colours which are much bluer than the typical passive E/S0 cluster
member mean that its nature is not immediately obvious from either the
published spectrum or the true colour images of the cluster (Ivison et
al.\ 2000, 2001).

How does this effect the interpretations of SMM\,J140104.96 by Downes
\& Solomon (2003) and Ivison et al.\ (2001)?  We confirm that J1 is a
foreground lens as stated by Downes \& Solomon (2003).  However, our
observations disprove their central claim that J1n, J1se, and J2 are
highly-amplified multiple images of an intrinsically low-luminosity
sub-mm source. In fact, J1n/J1se and J2 are probably single images of
three background galaxies at $z=2.56$ (see Tecza et al.\ 2004), with
the H$\alpha$ emission from this system arising entirely from very red
J1n/J1se -- which is also the site of a massive gas reservoir and hence
most likely the far-infrared source (Frayer et al.\ 2003; Ivison et
al.\ 2001).  The addition of J1 to the foreground lens model does
increase the estimate of the area-averaged amplification for J1n/Jse
from 2.5 to $\sim 5$; however, this does not significantly alter any of
the conclusions in Ivison et al.\ (2001).

\medskip

\noindent {\it SMM\,J163650.43+405734.5} --- We also observed SMM\,J163650.43 
(ELAIS-N2\,850.4, Smail et al.\ 2003) using NIRSPEC for a total of
4.8\,ks at two orthogonal position angles.  The detailed kinematics of
this complex merging system using three dimensional near-infrared
spectroscopy will be discussed in Swinbank et al.\ 2005.  However, we
note that the broad H$\alpha$ component is apparently offset from the
narrow H$\alpha$ component by $\sim 300\kms$.  The redshift quoted in
Table~2 is that of the narrow line H$\alpha$.

\subsection{H$\alpha$ Properties of SMGs}

We show near-infrared spectra of the 23 detected galaxies from our Keck
and VLT observations in Fig.~1 marked with the expected redshifts of
lines based on the best-fit H$\alpha$ redshift (Table~2).  As these
were short exposures, primarily meant to derive redshifts, the
signal-to-noise ratio on the individual galaxies is generally modest.
To overcome this, we have also combined all of the spectra to provide a
composite near-infrared spectrum for a representative far-infrared
luminous galaxy at $z\sim2.4$.

We create the composite spectrum by de-redshifting and summing all of
the spectra (normalised by H$\alpha$ flux) for our galaxies (we note
that stacking the spectra based on their individual signal-to-noise
ratio or an unweighted stack does not alter any of the conclusions
below).  We also derive a composite spectrum for those galaxies which
individually show no signs of an AGN (i.e.,\ those with small [N{\sc
  ii}]/H$\alpha$ ratios and line widths).  The resulting composite
spectra are shown in Fig.~6.  The rest-frame composite spectrum from
the entire sample is best fit ($>99\%$ confidence level) with an
underlying broad-line region with a narrow-to-broad line flux ratio of
$0.6\pm0.1$ and FWHM$_{\rm rest}$ of $1300\pm210\kms$ for the
broad-line H$\alpha$ and $325\pm30\kms$ for the narrow-line H$\alpha$.
The average [N{\sc ii}]/H$\alpha$ for the entire sample is
$0.42\pm0.05$ which is indicative of star formation.

%
%
\centerline{ \psfig{file=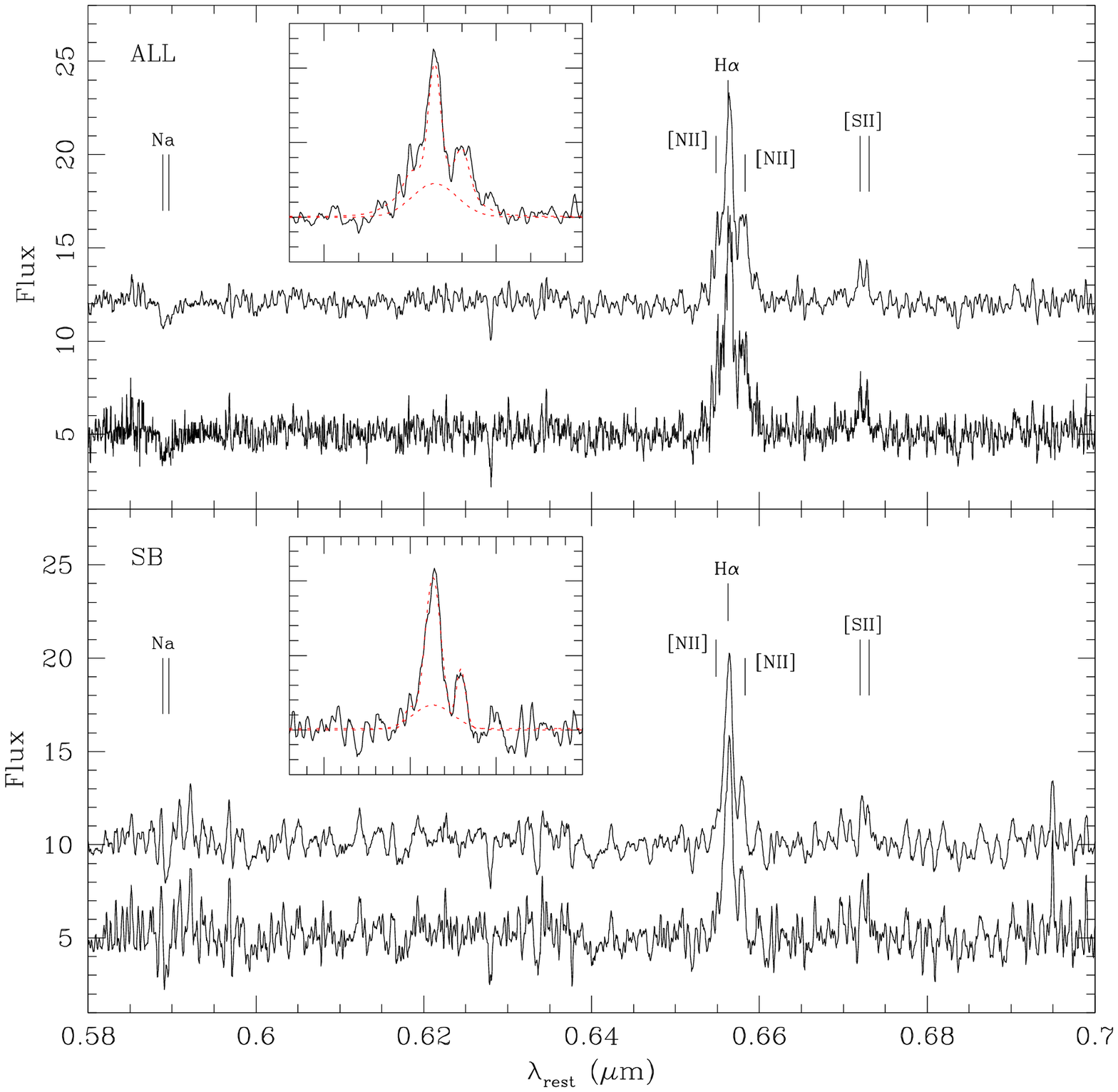,width=3.5in,angle=0}} {\noindent
  \footnotesize\addtolength{\baselineskip}{-2pt} {\sc Fig.~6. --}
  Rest-frame composite spectrum of all of the galaxies in our sample
  ({\it top}) and well as the composite from the (individually
  spectroscopically classified) star-forming galaxies ({\it bottom})
  (we have not included SMM\,J123707.21 in either composite as we
  believe that the spectra display two H$\alpha$ components coming from
  two interacting galaxies rather than H$\alpha$ and [N{\sc ii}]).  The
  insets show the region around the H$\alpha$ emission line with the
  best fit to the emission line using four Gaussian profiles overlaid.
  We also show the broad line component of the best-fit.  The average
  [N{\sc ii}]/H$\alpha$ ratio is $0.42\pm0.05$ for the entire sample
  and $0.19\pm0.05$ for the star-forming galaxies.  Both are consistent
  with star-forming galaxies rather than an AGN.  We also detect the
  [S{\sc ii}]$\lambda\lambda$6716,6731 emission doublet (marginally
  detected in the SB sample) and the (unresolved) stellar
  Na$\lambda\lambda$5889.95,5895.92 absorption doublet.  By fitting the
  Na absorption and [S{\sc ii}] doublets, we find no velocity offset
  from H${\alpha}$.  The [S{\sc ii}]/H$\alpha$ ratio can be used to
  classify the SMGs/ORFGs optical emission line properties as LINER- or
  H{\sc ii}-region-like.  While the SMGs/OFRGs which are individually
  classified as star-forming galaxies show no signs of an underlying
  broad H$\alpha$ line, the composite from this subsample of galaxies
  is best fit with a narrow/broad H$\alpha$ emission line ratio of
  $0.45\pm0.20$. }

\medskip

The [S{\sc ii}]/H$\alpha$ ratio can be used to classify the spectral
type of galaxies (Veilleux et al.\ 1995).  The wavelengths of the
[S{\sc ii}]$\lambda\lambda6716,6731$ lines in the composite spectra are
$6722\pm6$ and $6729\pm5$\AA\ -- indicating no detectable velocity
offset between the H$\alpha$ and [S{\sc ii}] emission lines. From the
strengths of the lines we estimate a ratio of [S{\sc ii}]/H$\alpha =
0.10\pm0.04$ -- placing the composite SMG/OFRG within either the LINER
or H{\sc ii} region of the classification space (Veilleux et al.\ 1987,
1995).  This is similar to the typical optical spectral classification
of local ULIRGs, for which mid-infrared observations (Lutz et al.\ 
1999) and spatially-resolved spectroscopy (Heckman et al.\ 1990) both
suggest that star-formation is the dominant power source (see Lutz et
al.\ 1999).

The composite spectrum also shows an absorption feature at a wavelength
close to that expected for the Na doublet which is seen from cool stars
and from the warm neutral (T$\sim10^{4}$\,K) phase of the interstellar
medium (ISM; Phillips 1993).  This has been used to map the velocity
structure of large-scale outflows from nearby ULIRGs (Heckman et al.\ 
2000; Rupke et al.\ 2002).  For a galaxy whose light is dominated by a
young star-burst, as we believe is the case for these SMGs/OFRGs, the
warm ISM is expected to be the primary source of this absorption
feature.  We fit the Na absorption line with two Gaussian profiles with
a fixed separation and variable width and derive a central wavelength
of 5889$\pm$5\AA, which is entirely consistent with the rest-frame
wavelength of the resolved doublet
($\lambda\lambda5889.95,5895.92$\AA).  We thus place a limit of
$<$150$\kms$ on the possible velocity offset of this feature from the
H$\alpha$ redshift.  The FWHM$_{\rm rest}$ of the Na lines is
$292\pm192\kms$, which is also consistent with the FWHM of the
H$\alpha$ emission in the rest-frame composite.

The composite spectrum thus shows no signs of an offset in velocity
between Na or [S{\sc ii}] and H$\alpha$.  This is slightly surprising
since many local luminous/ultra-luminous infra-red galaxies display
velocity offsets of several 100's km\,s$^{-1}$ between these lines
(Heckman et al.\ 2000; Rupke et al.\ 2002).  We defer a detailed
discussion of the velocity offsets between the UV ISM and (star-burst)
nebular emission lines in the individual galaxies to a later paper
(Smail et al.\ in prep).

Turning to the rest-frame composite from those galaxies which
individually show no signs of an AGN component (i.e.\ those with low
[N{\sc ii}]/H$\alpha$ emission line ratios and small H$\alpha$ line
widths), we fit this spectrum with both [N{\sc
  ii}]$\lambda\lambda$6548,6583 emission lines and also attempt to fit
a broad line H$\alpha$ component.  When the broad component is included
in the fit, the total $\chi^{2}$ is marginally better than when the
broad component is excluded, with a change in the total $\chi^{2}=4$
which corresponds to $\sim2\sigma$ (or $\sim87\%$ confidence).  The
resulting best-fit model for the SB composite has an underlying
broad-line H$\alpha$ component with a ratio of broad-line/narrow-line
H$\alpha=0.45\pm0.20$ and a broad-line FWHM$_{\rm
  rest}=890\pm210$\,km\,s$^{-1}$, suggesting that even the SMGs/OFRGs
which are identified individually as star-bursts may contain at least
some level of underlying non-thermal activity.  Nevertheless, the
[N{\sc ii}]/H$\alpha$ emission line ratio is $0.19\pm0.05$, which,
along with the limit of [S{\sc ii}]/H$\alpha \ls 0.14$, still suggests
that the energy output is star-formation-, rather than AGN- dominated.

\smallskip

\subsection{Kinematics of H$\alpha$ emission}

The narrow H$\alpha$ emission line of the SB composite has a rest-frame
FWHM$_{\rm rest}$, after correcting for the instrumental resolution of
the observations, of FWHM$_{\rm rest}=400\pm70$\,km\,s$^{-1}$ -- in
agreement with the average H$\alpha$ line width from the SMGs/OFRGs
which make up this composite spectrum ($350\pm50\kms$).  This is
somewhat larger than the H$\alpha$ line widths of UV-selected galaxies
at $z\sim 2$ identified by Erb et al.\ (2003), who derive a mean
FWHM$_{\rm rest}=242\pm65$\,km\,s$^{-1}$ for their sample.  We compare
the distribution of line widths for these two populations (as a
function of their H$\alpha$ luminosities) in Fig.~7.  We see that the
SMGs/OFRGs are typically 6 times brighter in H$\alpha$ ({\it before}
any correction for reddening) than the UV-selected population at their
epoch.  More interestingly, the emission line widths of the SMGs/OFRGs
are on average $\sim $\,50\% larger than those measured for the
UV-selected systems, although the two distributions overlap
substantially.  This difference could reflect either: (1) different
halo masses, (2) different dynamical states, (3) a contribution from a
broad AGN component in some SMGs/OFRGs, or (4) starburst-driven
superwinds contributing to the line widths. The composite spectra from
\S3.2 certainly suggest that an unidentified broad component to the
H$\alpha$ emission may be present in some SMGs/OFRGs -- however, as we
show below, we do not believe that this is the chief cause of the
difference in the line widths of the two populations.

%
%
\addtocounter{figure}{1}
\begin{figure*}[tbh]
\centerline{ \psfig{file=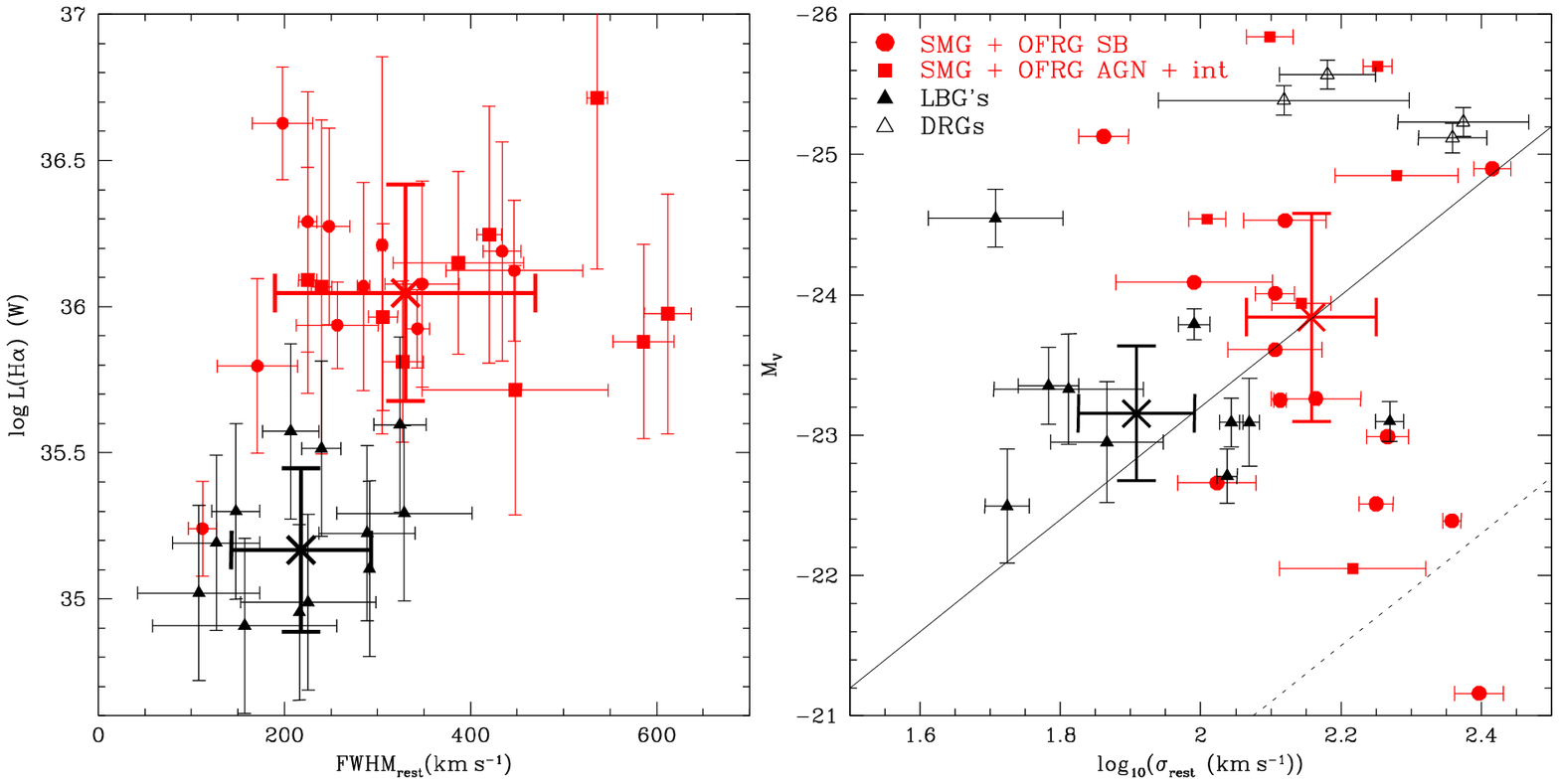,width=6.0in,angle=0}}
\caption{\footnotesize
  {\it Left:} Comparison of the H$\alpha$ luminosity vs FWHM$_{\rm
    rest}$ for the narrow H$\alpha$ components (deconvolved for
  instrumental resolution) in our sample compared to those found in
  UV-selected galaxies by Erb et al.\ (2003; we have conservatively
  assumed a factor of two uncertainty in the H$\alpha$ luminosities for
  this sample).  We also include those galaxies which show single broad
  lines (no narrow component). The crosses mark the positions of the
  median luminosity and FWHM$_{\rm rest}$ for the two populations.  The
  line widths in the SMGs/OFRGs have a median of $330\kms$ and a long
  tail out to $\sim600\kms$.  In comparison to the UV-selected galaxies
  studied by Erb et al.\ (2003) (which have a median of $\sim210\kms$),
  SMGs/OFRGs appear to be typically more massive systems, with a wider
  distribution of line widths (although some of the broadest lines may
  come from unresolved companions or AGN).  {\it Right:} Correlation
  between rest-frame optical emission line width and rest-frame
  $V$-band luminosity (the Faber--Jackson relation) for the SMGs/OFRGs
  compared to the LBGs and DRGs from Pettini et al.\ (2001) and van
  Dokkum et al.\ (2004).  The crosses mark the positions of the median
  luminosity and $\sigma_{\rm rest}$ for the SMGs/OFRGs and LBGs.  We
  show the local Faber--Jackson relation by the dotted line (assuming
  $V-R$=0.3) and also the same line offset by a further 2.5 magnitudes
  in order to pass through the median of the SMG sample (\it {solid
    line}). }
\end{figure*}

We first compare our H$\alpha$-derived line widths with those obtained
from the dynamics of cold gas as traced by interferometric maps of the
CO distributions in a small number of SMGs (Frayer et al.\ 1999; Neri
et al.\ 2003; Genzel et al.\ 2003).  The sample of five galaxies
compiled by Neri et al.\ (2003) have a FWHM$_{\rm rest}$ for the CO of
420$\pm$35\,km\,s$^{-1}$ and a mean ratio of H$\alpha$ to CO FWHM
consistent with unity ($1.27\pm 0.24$) for the four galaxies in common
with our sample.  The CO line widths are unaffected by the presence of
any AGN or large-scale winds, and so this suggests that the AGN or
superwind contribution to the H$\alpha$ line widths for SMGs/OFRGs may
not be responsible for the comparatively large H$\alpha$ line widths.

We can therefore combine the emission-line widths for our sample with
the typical physical extent of the H$\alpha$ emission from our
narrow-band imaging of the SMGs/OFRGs to place limits on their masses.
The spatial extent of the H$\alpha$ in the galaxies has a wide
distribution: $\ls 0.5$--1.0$''$ (corrected for seeing) or $\ls
4$--8\,kpc (c.f.\ Smail et al.\ 2004).  Assuming that the H$\alpha$
emission arises from virialised clouds in the galaxy's potential well,
we estimate a typical mass of 1--2$\times 10^{11}$\,M$_\odot$ for the
SMGs/OFRGs in our sample (Erb et al.\ 2003), with corresponding
dynamical times of 10--20\,Myr.  Using the limits on the spatial extent
of the CO emission, Neri et al.\ (2003) determine a median dynamical
mass of $\sim6\times10^{10}M_{\odot}$ assuming that the CO gas is in
bound orbits.  Thus, the masses derived from the dynamics of the cold
gas in a small sample of these galaxies support those estimated from
the emission-line kinematics.

The masses of UV-selected galaxies at $z\sim 2$ derived in an identical
manner by Erb et al.\ (2003) have a median of
$3.3\pm1.1\times10^{10}M_{\odot}$, around 5 times lower than our
estimates for the SMGs/OFRGs.  This is due to a combination of smaller
estimated sizes and lower H$\alpha$ line widths.  Clearly both of these
estimates have large systematic errors, yet they are suggestive of a
real difference in the characteristic masses (or dynamical states) of
restframe UV- and far-infrared selected galaxies at $z\gsim 2$.

We can also obtain an independent measure of the masses of the
SMGs/OFRGs from the seven galaxies which show multiple components and
for which we have estimates of the velocity differences between the
components. Assuming random orientations on the sky and that the
components are bound/merging, the masses of the systems are
$\sim1.5\pm0.9\times10^{11}M_{\odot}$.  This is comparable to the
earlier estimates and gives us confidence that the SMGs/OFRGs are indeed
massive galaxies.

\subsection{Magnitude -- Line Width correlations}

Correlations between rest-frame luminosity and kinematics (as measured
by the widths of the emission lines) at high redshift have had only
limited success.  Pettini et al.\ (2001) and van Dokkum et al.\ (2004)
have attempted to study the Tully-Fisher-like or Faber-Jackson-like
(Tully \& Fisher 1977; Faber \& Jackson 1976) correlations for LBGs and
distant red galaxies (DRGs) at z$\gsim$2; however, no correlations have
been found over the range FWHM$_{\rm rest}\,\sim 120 - 300\kms$.  This
may be due in part to the small sample sizes involved.  To search for
such a correlation, the SMGs for which we have well-defined narrow-line
H$\alpha$ line width measurements can be added to this sample.  We
compute the de-reddened rest-frame $V$-band magnitudes for the SMGs in
our sample by using {\sc hyper-z} (Bolzonella et al.\ 2000) to compute
the best-fit SED to the observed $IJK$ photometry from Smail et al.\ 
(2004).  We also convert FWHM$_{\rm rest}$ to $\sigma$ by assuming
FWHM$=2.35\times\sigma$.

Fig. 7 shows the resulting luminosity--line width correlation for the
combined sample of SMGs, LBGs, and DRGs (the latter two from Pettini et
al.\ (2001) and van Dokkum et al.\ (2004) are corrected for reddening
using their estimates of $A$$_{V}$).  We also overlay the local
Faber--Jackson relation from J{\o}rgensen et al. (1995, 1999; assuming
$V-R=0.3$), and fit the zero-point of the same correlation so that it
passes through the median of the SMG sample -- resulting in an offset
of 2.5 mag.  This offset is comparable to the $\Delta_{\rm fade}V$
found in Smail et al.\ (2004) and suggests that the descendants of
these high redshift populations are likely to lie on or around the
local Faber--Jackson relation.

\subsection{SFR Comparisons}

Next we compare the far-infrared and H$\alpha$ luminosities of the
galaxies in our sample to investigate the influence of AGN and star
formation power-sources and the possible effects of dust extinction on
the H$\alpha$ emission from this population.

Fig.~8 compares the H$\alpha$ and far-infrared luminosities of the
galaxies in our sample. We have also included measurements from the
literature from five well-studied dusty, high-redshift galaxies:
SMM\,J02399$-$0136 (Ivison et al.\ 1998, 2000), SMM\,J17142+5016 (Smail
et al.\ 2003), ERO\,J164502+4626 (Dey et al.\ 1999), SMM\,J04431+0210
(Frayer et al.\ 2003), and SMM\,J16359+6612 (Kneib et al.\ 2004).

The star formation rate derived from the far-infrared and H$\alpha$
should be correlated if the effects of dust and any contributions from
AGN are uniform across the sample.  The SMGs/OFRGs in our sample show
only a weak correlation between SFR(FIR) and SFR(H$\alpha$), with
comparable scatter in the H$\alpha$-derived SFR and that estimated from
the far-infrared ($\Delta$(FIR)/FIR$=0.30\pm0.18$ vs.
$\Delta$(H$\alpha$)/H$\alpha=0.37\pm0.20$).  Reassuringly, the two
galaxies for which we failed to obtain H$\alpha$ detections in good
conditions are also two of the least luminous galaxies when ranked on
their far-infrared emission.

%
%
\begin{figure*}[tbh]
\centerline{\psfig{file=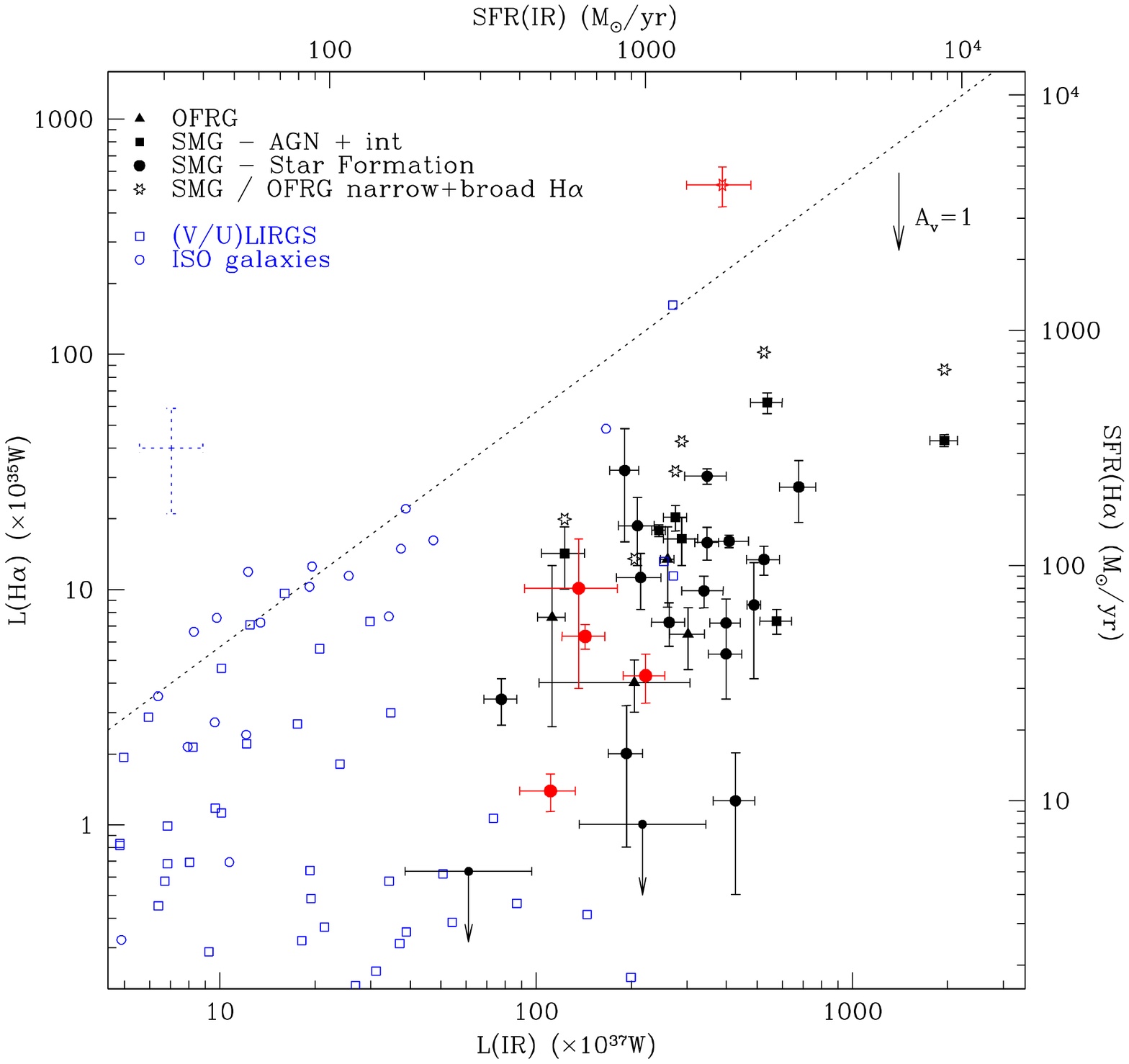,angle=0,width=6.0in}}
\caption{\small
  Comparison of the far-infrared vs. narrow-line H$\alpha$ luminosities
  and star formation rates in our data with those in local samples.  We
  include in the plot the SFRs for the SMGs/OFRGs when the broad-line
  H$\alpha$ flux is included in the estimate (the galaxies classified
  as intermediate in \S2.4 are included with the AGN classification).
  This is more comparable to the quantities calculated for local
  samples.  We also include the previously published SMGs (Kneib et
  al.\ 2004; Dey et al.\ 1999; Smail et al.\ 2003; Frayer et al.\ 2003;
  Ivison et al.\ 1998) as red points ordered from lowest to highest
  H$\alpha$ luminosity respectively.  The dotted line represents
  identical estimated SFR from H$\alpha$ and far-infrared.  The sample
  is compared to the results from ISOCAM, including the $z=0.2$--1.5
  galaxies in the Hubble Deep Field-South by Franceschini et al.\ 
  (2003) and the local luminous infra-red galaxy sample ($z=0.2$--0.7)
  by Flores et al.\ (2004).  We also compare the data to the Very
  Luminous and Ultra-Luminous infrared galaxy sample of local {\it
    IRAS} greater than 2Jy sources by Dopita et al.\ (2002) and
  Poggianti \& Wu (2000).  We include a representative error bar from
  the comparison samples on the left side of the plot.}
\end{figure*}

We have also included in Fig.~8 samples of similar and less luminous
far-infrared galaxies from surveys of the $z<1$ universe.  These come
from ISOCAM-selected galaxy surveys (Franceschini et al.\ 2003; Flores
et al.\ 2004) and studies of local Very- and Ultra-Luminous infrared
galaxies by Poggianti \& Wu (2000) and Dopita et al.\ (2002).  Compared
to these samples, we see a similar wide dispersion in the ratio of
far-infrared to H$\alpha$ luminosities across a factor of nearly 100 in
far-infrared luminosity.  This is suggestive of a similar diversity in
the energy sources and obscuration for galaxies with far-infrared
luminosities from 10$^{10}$--$\gsim$\,10$^{12}$\,L$_\odot$.  In support
of this, we note that the distribution of H$\alpha$ equivalent widths
(EW, Table~2) for the SMG sample peaks at about $\sim 20$--40\AA\ and
has a long tail out to $\gsim100$\AA, with a median
EW(H$\alpha$)\,$=75\pm 25$\AA.  The shape of the distribution is very
similar to that seen for the rest-frame H$\alpha$ EWs of local ULIRGs
from Veilleux et al.\ (1999), which have a median
EW(H$\alpha$)\,$=73\pm 8$\AA\, suggesting that the unobscured/partially
observed mix of emission-line gas and stellar continuum is comparable
to local ULIRGs (although we note that the projected size of the
spectroscopy apertures in the local ULIRGs is less than 1\,kpc, whereas
for the SMGs/OFRGs the projected size is $\sim10$\,kpc; thus some
caution should be taken when comparing the two samples).

Our SMG/OFRG sample is intrinsically luminous in the far-infrared, but
their H$\alpha$ flux (and estimated SFR) is much less than expected
with a median SFR(H$\alpha$) compared to SFR(IR) of $94\pm20$ and
$1380\pm190$M$_{\rm \odot}$yr$^{-1}$, respectively.  Overall, the
H$\alpha$ SFRs are suppressed by at least a factor of ten relative to
that suggested by the far-infrared.  In comparison to other
far-infrared-selected samples, we see that the SMGs/OFRGs extend the
trend for proportionally less H$\alpha$ luminosity in more far-infrared
luminous galaxies.  As we have shown there is no detectable difference
between the far-infrared luminosities of the starburst- and AGN-classed
galaxies in our sample -- suggesting that this declining ratio of
H$\alpha$ to far-infrared emission is unlikely to be caused by an
increasing AGN contribution.  Hence, we attribute the variation to
copious and increasing amounts of dust enshrouding the galaxy and
extinguishing H$\alpha$ emission.

Unfortunately, the wavelength coverage of our spectroscopic
observations does not extend to H$\beta$, and thus the reddening in
these galaxies cannot be estimated directly from the Balmer decrement
to confirm this suggestion, although attempts at estimating the
reddening in this manner are underway for a subsample of galaxies.  The
only target with an H$\beta$ measurement in the literature is
SMM\,J123707.21+621408.1 (Simpson et al.\ 2004), with an H$\beta$ flux
of $2.1\pm0.9\times10^{-20}\Wm2$.  The H$\alpha$/H$\beta$ flux ratio is
8$\pm$6, which corresponds to a reddening of $A_V=1.4\pm 1.0$ (Calzetti
et al.\ 1994); the large uncertainties arise from the modest
signal-to-noise ratio detections of both H$\alpha$ and H$\beta$
emission lines.  For the rest of the sample, we have attempted to
derive the reddening for these galaxies from their broadband
optical/near-infrared colours (Smail et al.\ 2004).  These are derived
by using the {\sc hyper-z} photometric redshift code (Bolzonella et
al.\ 2000) to fit young continuous star formation models with variable
reddening and age to the galaxy photometry at their known redshifts.
We find a median reddening of $A_V=3.0\pm1.0$.  Accounting for the
contribution of H$\alpha$ to the $K$-band photometry may decrease this
estimate slightly but the reddening estimated from the continuum
colours indicates substantial extinction at the wavelength of H$\alpha$
(assuming that the continuum and line emission arise from the same
regions in the galaxy). The scatter of a factor of 2.5 is more than
sufficient to explain the dispersion in the strength of the H$\alpha$
emission at a fixed far-infrared luminosity.  We conclude therefore
that the large scatter in SFR(FIR)/SFR(H$\alpha$) probably arises from
two main factors: (1) there is a range in the continuum extinction in
the more far-infrared luminous population and, (2) the morphological
diversity of sub-mm-selected galaxies -- which includes a large
fraction of interacting or merging systems, sometimes with
highly-obscured components may lead to a large variation in the in-slit
H$\alpha$ fluxes for these systems (Chapman et al.\ 2003b; Smail et
al.\ 2004).

\subsection{Metallicities}

Rest-frame optical emission lines from extra-galactic H{\sc ii} regions
provide an important diagnostic of the chemical evolution of galaxies
since their properties reflect the make-up of the (ISM).  Moreover,
feedback processes from short-lived massive stars in high SFR galaxies
may be responsible for the enrichment of the intergalactic medium (IGM)
with heavy metals over a wide range of redshifts (Heckman et al.\ 2000;
Adelberger et al.\ 2003).  For star-forming and irregular galaxies the
correlation between metallicity (traced by Oxygen abundances, O/H) and
rest-frame luminosity is well established in the local universe and
spans a factor of over 100 in (O/H) and at least eight magnitudes in
$M$$_{V}$.  The correlation is in the sense that more massive (and
luminous) galaxies exhibit a higher degree of metal enrichment (Garnett
et al.\ 1987; Zaritsky et al.\ 1994; Garnett 2002; Lilly et al.\ 2003;
Lamareille et al.\ 2004).  If such a correlation exists at high
redshift then it may be more important for understanding the present
distribution of metals in the universe, since the bulk of the
star-formation activity in the most massive and active galaxies is
thought to occur at $z>1$ (Blain et al.\ 1999b).

%
%
\begin{figure*}[tbh]
  \centerline{\psfig{file=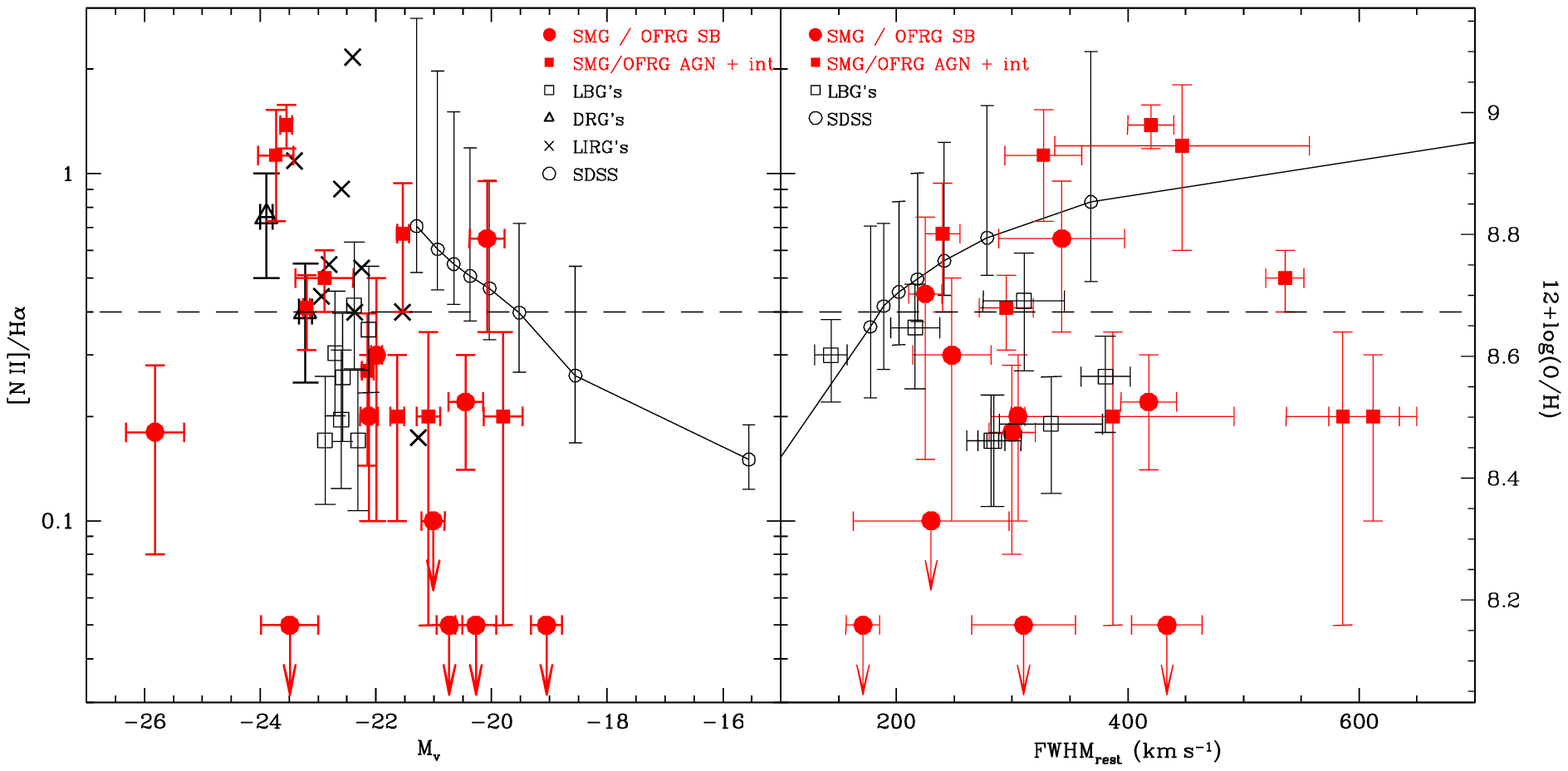,angle=0,width=6.0in}}
\caption {{\it Left:} Metallicity-Luminosity relation for SMGs/OFRGs. 
  {\it Right:} Line width - metallicity relation for the SMGs/OFRGs.
  The abundance is derived using the N2 index calibration given by
  Pettini \& Pagel (2004).  We show two low-redshift samples for
  comparison, one from the SDSS DR2 (Abazajian et al.\ 2004) and one
  for local luminous infrared galaxies (Armus et al.\ 1989).  For a
  higher redshift comparison sample, we show the metallicities for the
  $z\sim 2$ UV-selected galaxies from Shapley et al.\ (2004) and for a
  near-infrared-selected sample of luminous, $z\sim 2$ galaxies from
  van Dokkum et al.\ (2004).  The horizontal dashed line represents
  solar metallicity.  The SMGs/OFRGs exhibit a large range in $M_{V}$
  and FWHM$_{\rm rest}$, and many have slightly sub-solar or solar
  abundances.  Note that these are the observed $V$-band luminosities
  (uncorrected for reddening; c.f.\ Fig.~7).}
\end{figure*}

Since we lack the observations necessary to calculate the $R_{23}$
index (Zaritsky et al.\ 1994) for our sample, we turn to the $N2$ index
described by Storchi-Bergmann et al.\ (1994) and discussed more
recently by Denicolo et al.\ (2002) and Pettini \& Pagel (2004).  This
indicator is defined at $N2=log$([N{\sc ii}]$\lambda$6584/H$\alpha$)
and is calibrated to the oxygen abundance (O/H) via
$12+log($O/H$)=8.90+0.57\times N2$ (Pettini \& Pagel 2004).  However,
this calibration remains uncertain, and there are obvious drawbacks to
using this index for young galaxies in which AGN may play a role in
defining the emission-line characteristics and in which primary and
secondary sources of Nitrogen production may lead to mismatches in N/O.
For this reason we have chosen to compare populations directly using
their $N2$ measurements (as opposed to (O/H) measured from a range of
indicators) to minimise systematic uncertainties arising from the
calibration from $N2$ to (O/H). We therefore determine the $N2$ index
for all of the galaxies in our sample and give upper limits where the
[N{\sc ii}] emission line is not detected with sufficient significance
to measure accurately.

In Fig.~9 we show the distribution of $M$$_V$ versus $N2$ for
SMGs/OFRGs as a proxy for their Luminosity--Metallicity relation and
compare this with similar observations of both local and high redshift
galaxy populations.  To calculate the observed rest-frame $V$-band
luminosities of the SMGs we use the photometry from Smail et al.\ 
(2004) and fit SED's based on their observed broad-band $IJK$
magnitudes, which, at $z\sim2.4$, corresponds to rest-frame $UBR$,
bracketing the $V$-band.  As Smail et al.\ (2004) show, the competing
effects of dust reddening and fading of the young stellar populations
almost cancel each other ($\Delta_{\rm dust}V\sim 3$--4, $\Delta_{\rm
  fade}V\sim 4$ based on fitting the broadband colors of this
population). Thus, we expect that the present-day descendants of these
galaxies would have absolute luminosities not very dissimilar to those
we have estimated (see also Fig.~7).  For the comparison samples we
convert the rest-frame $B$-band magnitudes for the UV-selected $z\sim
2$ sample from Shapley et al.\ (2004) using their $(R-K)$ colors to
predict the rest-frame $V$-band magnitudes.  We also convert the
median, 10 and 90 percentile trend lines from a volume-limited
($z<0.1$) local emission-line galaxy sample from the SDSS DR2
(Abazajian et al.\ 2004; although we stress that the Sloan aperture
only samples the central $\sim3$\,kpc of a galaxy at $z=0.05$ whereas
the NIRSPEC slit will sample $\sim10$\,kpc at $z=2.3$).  We see at
least an order-of-magnitude range in $N2$ for the SMGs/OFRGs
(discounting obvious AGN), with little correlation between the line
index and rest-frame luminosity.  The implied median metallicity is
slightly below solar, and appears to be similar to that inferred for
bright, UV-selected galaxies at $z\sim 2$ (Shapley et al.\ 2004).

To address the competing and uncertain effects of dust reddening and
passive fading of the stellar populations on the $M$$_V$ versus [N{\sc
  ii}]/H$\alpha$ plot, we have also constructed a FWHM$_{\rm H\alpha}$
versus [N{\sc ii}]/H$\alpha$ diagram, adopting the FWHM$_{\rm H\alpha}$
as a crude proxy for the dynamical mass of the galaxies.  As can be
seen, the large dynamical masses we inferred for the SMG/OFRG
population suggest that their present-day descendants are likely to be
luminous and metal-rich (super-solar) systems (as shown by the trends
seen in the SDSS dataset plotted in Fig.~9). The apparently modest
metallicities we measure would then indicate that these systems are
seen during an early phase of enrichment -- suggesting that they are
relatively youthful galaxies and arguing against them undergoing a
cycle of repeated short ($\sim 10$\,Myr) bursts of star-formation over
a relatively extended period ($\gsim 1$\,Gyr; Smail et al.\ 2003).
However, there are other possible explanations for the apparently low
metallicities of these galaxies.  First, we note that star-formation
can usually only act to increase a galaxy's metallicity (and therefore
$N2$), and in particular starburst-driven feedback mechanisms are
unlikely to preferentially expel large quantities of metals without
entraining and expelling associated gas.  Only an infall of unenriched
material into the galaxy could cause the metallicity and effective
heavy-element yield to decrease (e.g.,\ Garnett 2002).  Alternatively,
if these systems are very young and the halos of the SMGs/OFRGs have
yet to coalesce (as suggested by the clear merger/interacting
morphologies of many systems), then the metallicities we measure may
reflect those of the progenitor components.  The similarity of the $N2$
estimates with those for the UV-selected population at this epoch could
then be interpreted as indicating that the SMGs/OFRGs arise from
mergers among the UV population.  The slightly more evolved descendants
of these mergers will be able to retain their enriched gas and so
produce a super-solar stellar population.  We stress that local
calibrations of the $N2$ index from H{\sc ii} regions can show
super-solar metallicities which may not reflect the global abundance in
a galaxy.  However, since our spectroscopic slit covers most of the
area of our galaxies and the measured H$\alpha$ fluxes suggest
substantial star-formation, we suggest that the $N2$ index should
provide a fair estimate of the metallicity of the gas in the galaxy as
a whole.  We caution that the nitrogen abundances that we measure come
from primary metal enrichment, whereas the local N/O versus O/H
relations suggest that for metal-rich systems the nitrogen is secondary
in origin (produced from intermediate mass stars far removed from the
first generation).  The evolution of the nitrogen
metallicity--luminosity relation might be therefore quite different
from that shown for oxygen or other $\alpha$-elements.  We also note
that the $N2$ index is not an ideal metallicity indicator given the
presence of an AGN in many of these galaxies, and we look forward to
future studies based on [O{\sc ii}], [O{\sc iii}] and H$\beta$ emission
lines.

\subsection{X-ray comparisons}
Using deep {\it Chandra} observations of the Hubble Deep Field (HDF),
ELAIS N2 field, and SSA13 field (Alexander et al.\ 2003b; Manners et
al.\ 2003; Mushotzky et al.\ 2000), it is possible to compare the X-ray
and H$\alpha$ properties of the SMGs/OFRGs which overlap with the {\it
  Chandra} coverage.  Of the 18 sources which were covered by {\it
  Chandra}, nine were detected in the 2--8\,keV hard X-ray band, mostly
in the HDF owing to the much deeper X-ray observations available for
that field.  We convert the observed 2--8\,keV flux to a rest-frame
2--10\,keV luminosity using $L_{X} = 4\pi
d_{L}^{2}f_{X}(1+z)^{\Gamma-2}$, which takes into account the
$k$-correction (Alexander et al.\ 2003b), assuming a spectral index
$\Gamma=2$.

In Fig~10 we compare the H$\alpha$ luminosities from the SMGs/OFRGs to
their X-ray luminosities and contrast these with a local sample of {\it
  IRAS}--selected Seyfert galaxies analysed by Ward et al.\ (1988).  We
also overlay two lines showing the correlation of $L$$_{(2-10{\rm
    KeV})}$ versus $L$(FIR) from Ranalli et al.\ (2003). The first line
simply converts their $L$(FIR) to $L$(H$\alpha$) assuming that the
relation in \S2 holds, and the second line assumes a further 10 times
suppression of the H$\alpha$ emission (relative to the far-infrared) as
indicated by Fig.~8.

The distribution of the SMGs/OFRGs in Fig.~10 roughly follows that seen
for the Ward et al.\ (1988) sample -- {\it irrespective} of the
spectral classification of the SMGs/OFRGs.  However, we note that the
correlation for local star-forming galaxies from Ranalli et al.\ (2003)
-- if scaled for the relative underluminosity of H$\alpha$ in the
SMGs/OFRGs -- can explain the properties of the least X-ray luminous
galaxies, these comprise roughly half of our sample.

%
%
\centerline{ \psfig{file=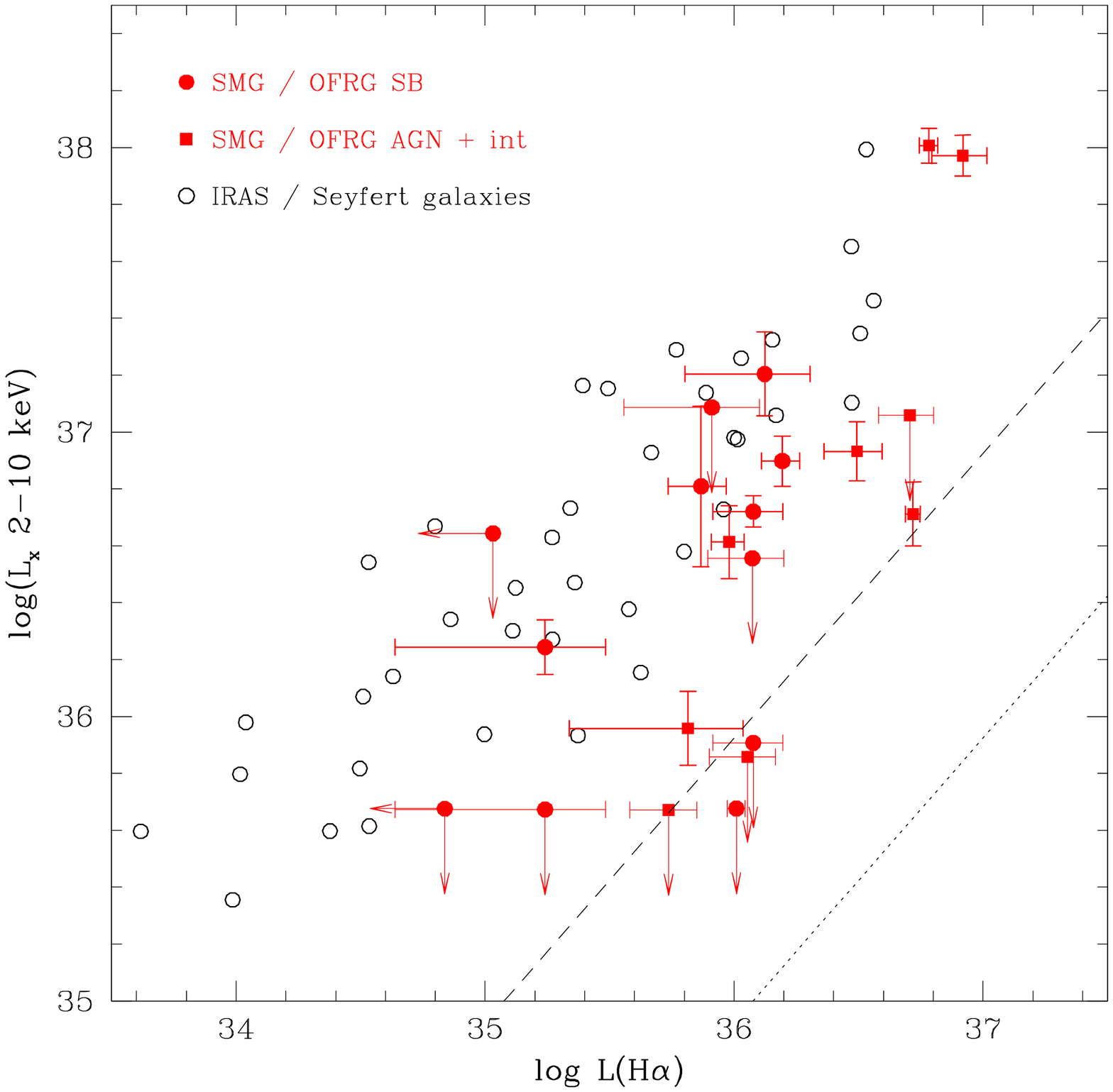,width=3.5in,angle=0}} {\noindent
  \footnotesize\addtolength{\baselineskip}{-2pt} {\sc fig.~10. --}
  Distribution of H$\alpha$ and X-ray luminosities for the SMGs/OFRGs
  in our sample as compared to the local {\it IRAS}-selected Seyfert
  galaxies from Ward et al.\ (1988).  For the SMGs/OFRGs sample we plot
  the total H$\alpha$ luminosity from both narrow and broad components
  (if present).  We also compare the data to a sample of nearby
  star-forming galaxies by Ranalli et al.\ (2003), shown by a dotted
  line (converting their $L$(FIR) to $L$(H$\alpha)$ assuming the
  correlation in \S2 holds) and include the same line but assuming a 10
  times suppression of the H$\alpha$ emission (relative to the
  far-infrared) as indicated in Fig.~8, shown by a dashed line.  The
  plot shows that some SMGs/OFRGs which are identified as star-forming
  in the near-infrared (with FWHM $\leq 500\kms$ and [N{\sc
    ii}]/H$\alpha\lsim0.7$) show 2--10\,keV X-ray fluxes consistent
  with AGN luminosities.  It is also interesting to note that some of
  the SMGs/OFRGs which are spectroscopically-classified as AGN from
  their near-infrared spectra are not detected in the hard X-ray band.}

\medskip

The most intriguing galaxies in Fig.~10 are those which have high X-ray
luminosities ($\log L_{x}\gsim36.5$) but have restframe optical
spectral classifications of star-formation -- the X-ray luminosity in
these galaxies probably arises from an AGN which is so highly obscured
that either it is hidden in the rest-frame optical spectra (given their
modest signal to noise) or we missed the AGN component with our
spectroscopic slit in these systems.  The presence of a broad-line
component in the composite star-forming spectrum in \S3.2 would support
the former interpretation.

We also point out that there are six SMGs/OFRGs which are not detected
to a flux limit of $1.4\times10^{-19}\Wm2$ in the 2-Ms {\it Chandra}
observations of the HDF (two) or $2.2\times10^{-18}\Wm2$ in the ELAIS
N2 (four) field.  While three of these are classified as SB from their
restframe optical spectra, three others show some signs of AGN in the
near-infrared spectra (with broad H$\alpha$ or large [N{\sc
  ii}]/H$\alpha$ ratios).  The X-ray limits on two of these
(SMM\,J163650.43 and SMM\,J163706.51) are particularly stringent -- yet
they show clear AGN signatures in their near-infrared spectra.  The
detection of broad H$\alpha$ and lack of X-ray emission are difficult
to understand -- with these galaxies being almost an order of magnitude
fainter in the X-ray waveband than expected from their H$\alpha$
luminosities.  Given the amount of gas available to fuel an AGN in
these systems, we conclude that they are likely to be intrinsically low
luminosity and probably have low mass central black holes.

\section{Conclusions}

We present the results of near-infrared spectroscopic and narrow-band
detections or limits on the H$\alpha$ emission from a sample of 30
ultraluminous, dusty galaxies at $z\sim1.4$--2.7.  The majority of
these galaxies come from sub-mm/mm-surveys, with a small number
identified as probable hot, luminous far-infrared sources from their
radio emission.  We see no difference between the properties of these
two samples in any of our diagnostic diagrams -- supporting the claimed
similarity of the two populations (Chapman et al.\ 2004b).

We identify the H$\alpha$ emission in the near-infrared spectra and use
the spectra to classify AGN by flagging galaxies with large [N{\sc
  ii}]/H$\alpha$ ratios ($\geq0.7)$ and/or large H$\alpha$ FWHM
($>500\kms$).  We find that the ratio of AGN- to star formation-
dominated galaxies from the rest-frame optical spectroscopy is roughly
40\%:60\% -- similar to the proportions estimated from their UV spectra
by Chapman et al.\ (2003a, 2003b, 2004c).  By constructing a rest-frame
composite spectrum for the entire sample, we find that average [N{\sc
  ii}]/H$\alpha$ ratio is $0.42\pm0.05$, which suggests that the
composite is star-formation- rather than AGN- dominated, although it
has an underlying broad-line component.  Furthermore the composite
spectra show both NaD absorption and [S{\sc ii}] emission features,
although we find no evidence of velocity offsets between these features
and the H$\alpha$ emission.  The [S{\sc ii}]/H$\alpha$ emission ratio
in the composite spectrum is $0.10\pm0.04$ -- indicating that the
spectral properties of our sample are comparable to a LINER or H{\sc
  ii} region -- similar results are found from the spectral
classifications of local ULIRGs (Veilleux et al.\ 1987, 1995).

We also derive the composite spectrum for those galaxies which,
individually, show no signs of an AGN in their near-infrared spectra.
This composite has an average [N{\sc ii}]/H$\alpha$ flux ratio of
$0.19\pm0.05$ and FWHM$_{\rm rest}$ of $400\pm70\kms$.  However, the
most striking result is that the composite spectrum appears to show an
underlying broad H$\alpha$ line with a broad/narrow H$\alpha$ flux
ratio of $0.45\pm0.20$, suggesting that even these galaxies may host a
low-luminosity AGN which is undetectable in our modest signal-to-noise
spectroscopy.

In seven of the systems with spectroscopic observations we find
velocity structure in the H$\alpha$ emission line. By comparison with
high-resolution broadband imaging, we identify these galaxies as
multi-component (probably interacting) systems with typical velocity
offsets between components of 100--600$\kms$.  This is not surprising
since locally many far-infrared luminous galaxies appear to be
disturbed/interacting systems.  Assuming that these are merging systems
with random orientations of their orbits on the sky, we estimate a
typical mass of $1.5\pm0.9\times10^{11}$M$_{\rm \odot}$.  We obtain a
similar estimate from the H$\alpha$ line widths of the whole sample.
These estimates are comparable to the dynamical mass estimates from CO
observations of a subset of these systems.

In all of the galaxies, we have attempted to deconvolve any broad
component to the H$\alpha$ line (which comes from an accretion disk
around a central super-massive black hole) from the narrow-line
H$\alpha$ flux (which comes from the star-forming regions).  Using the
narrow-line H$\alpha$ flux, we compare the SFRs of the SMGs/OFRGs as
compared to the SFR derived from the far-infrared emission.  The
SFR(H$\alpha$) versus SFR(FIR) correlation shows a large scatter, with
the SFR(H$\alpha$) typically a factor of ten less than what we would
expect from their far-infrared luminosities.  Most of this scatter,
however, can be explained by the reddening in these systems (estimated
from their broad-band photometry in Smail et al.\ (2004)).  The
suppression of the H$\alpha$ flux is therefore attributed to both
heavily obscured galaxies and a diverse range of morphologies.  The
average SFR derived from H$\alpha$ for the SMGs/OFRGs in our sample is
$94\pm20$M$_{\rm \odot}$yr$^{-1}$ (uncorrected for extinction).  Since
the continuum extinction correction at 6563\AA\ is $\sim$2.5
magnitudes, the total inferred SFR of these galaxies is expected to be
$\sim$1000M$_{\rm \odot}$yr$^{-1}$ -- comparable to that seen in the
far-infrared.  We also compare the SFR properties for our high-redshift
sample to local Very/Ultra-Luminous galaxy samples and find that the
scatter within our high redshift SMGs/OFRGs and the distribution of
equivalent widths are comparable to those of local Ultra/Very luminous
galaxy samples.  This suggests a similar range of obscured/unobscured
activity in the distant population to that seen locally, although with
a higher proportion of the star formation completely obscured from
view.

Using the $N2$ index, we have investigated the chemical abundances of
these galaxies and find that the $N2$ indices for the SMGs/OFRGs
suggest that they have slightly sub-solar metallicities, similar to
recent results from UV selected galaxies at these early epochs (Shapley
et al.\ 2004).  However, we note that the $N2$ indicator may not be a
reliable metallicity indicator for this population -- especially in the
presence of an AGN.  We find that the SMGs/OFRGs in our sample display
a large range in [N{\sc ii}]/H$\alpha$ versus $M$$_{V}$ or FWHM$_{\rm
  rest}$ (which we use as a proxy for their dynamical masses).

For the galaxies which are in the HDF, ELAIS N2 field and SSA13 field
we use their X-ray properties in order to further classify the
galaxies.  The SMGs/OFRGs classified as AGN on the basis of their
near-infrared spectra broadly follow the correlation seen between X-ray
and H$\alpha$ luminosities for local Seyfert 2 galaxies (Ward et al.\ 
1988).  We find that a subset of galaxies which are
spectroscopically-classified as star-forming in the near-infrared have
high X-ray luminosities, suggesting that they host highly obscured AGN.
Likewise, a small number of near-infrared spectroscopically-classified
AGN are undetected in deep {\it Chandra} observations.  We conclude
that these galaxies are likely to be intrinsically low luminosity and
probably have low-mass central black holes.

With observations of the H$\alpha$ emission from these SMGs/OFRGs at
$z\sim 2$ we can at last start to directly compare these galaxies to
similarly distant UV-selected systems.  We find that as expected the
SFR(H$\alpha$) for our sample is nearly an order of magnitude higher
than that found in $z\sim2$ UV-selected galaxies by Erb et al.\ (2003)
(who find an average SFR(H$\alpha$) of $21\pm3$M$_{\rm \odot}$yr$^{-1}$
-- even after their sample is corrected for extinction).  Similarly,
the line widths and dynamical information suggest that the halos of a
typical SMG/OFRGs may be upto 5 times more massive than the UV-selected
population.  We also find a higher rate of AGN activity in the
SMG/OFRGs -- suggesting the presence of actively fueled and growing
super-massive black holes in these galaxies.  However, somewhat
surprisingly, we find similar metallicities for the UV-selected and
more massive far-infrared luminous populations.  We suggest that this
may be explained if the SMG/OFRGs are relatively youthful, with their
deepening potential wells not yet sufficiently organised to retain a
larger fraction of the enriched material from their star formation
activity.

Overall our observations suggest that the high redshift SMG/OFRG
population shares many of the characteristics of similar (but somewhat
less luminous) far-infrared galaxies identified in the local Universe.
This includes the H$\alpha$ equivalent widths, the proportion of
obvious AGN and the typical optical spectral classification.  Previous
work has demonstrated the preponderance of merger-like morphologies in
the two populations and the similarity of their restframe optical
luminosities.  Yet there remain differences, with proportionally more
highly-obscured activity in the high-redshift population, apparently
larger dynamical mass, lower metallicities and higher gas fractions on
10-kpc scales.

We conclude that the SMG/OFRGs in our sample represent a population of
young, massive merging/interacting systems, the results of which cause
high instantaneous bursts of (highly obscured) star-formation and
actively fueled AGN activity.  Although these bursts are brief, they
can form all of the stars in an L$^\ast$ galaxy and in doing so will
raise the metallicity of these systems closer to that required by
observations of their likely present-day descendants: luminous
elliptical galaxies.

\acknowledgements We thank the referee for his constructive comments
and suggestions which significantly improved the content of this paper.
We would like to thank David Gilbank for adapting {\sc i-pipe} and
extensive help reducing the IRTF narrow-band imaging and Mike Balogh,
Chris Miller and Bob Nichol for providing the SDSS catalogs in a usable
format.  We acknowledge useful conversations or help from Dave
Alexander, Carlton Baugh, Richard Bower, Chris Done, Dave Gilbank,
Cedric Lacey, Max Pettini, Alice Shapley, Chris Simpson, Kaz Sekiguchi,
Tadafumi Takata and Richard Whitaker.  AMS acknowledges support from
PPARC, IRS acknowledges support from the Royal Society.  AWB
acknowledges support from NSF AST-0205937 and the Alfred Sloan
Foundation.

\end{document}